\documentclass[format=acmsmall, nonacm, screen=True]{acmart}
\setcopyright{none}
\AtBeginDocument{%
  \providecommand\BibTeX{{%
    \normalfont B\kern-0.5em{\scshape i\kern-0.25em b}\kern-0.8em\TeX}}}

\usepackage{enumitem}
\usepackage{longtable}
\usepackage{adjustbox}

\usepackage{multibib}
\newcites{nu}{Reference for supplementary literature}
\newcites{th}{References}

\setcitestyle{authoryear}

\begin{document}

\title{Bridging Level-K to Nash Equilibrium}
\subtitle{Accepted at \href{https://doi.org/10.1162/rest_a_00990}{\textit{The Review of Economics and Statistics}}}

\author{Dan Levin}
\authornotemark[1]
\authornotemark[2]

\author{Luyao Zhang}
\orcid{0000-0002-1183-2254}

\authornote{Corresponding authors: Luyao Zhang (email:lz183@duke.edu, institution: Data Science Research Center and Social Science Division, Duke Kunshan University, address: No.8 Duke Ave. Kunshan, Jiangsu 215316, China.) and Dan Levin (email: levin.36@osu.edu, institution: Ohio State University, address:433B Arps Hall, 1945 N. High St, Columbus, Ohio, 43210 USA }
\authornote{\textbf{Acknowledgements}: We thank the editor and two anonymous referees for valuable comments. We have benefited from the comments of participants in the Texas Experimental Economics
Symposium, Midwest Economic Theory, and Trade Conference, D-TEA workshop and from conversations with Yaron Azrieli, Pierpaolo Battigali, Tobias Brünner, Paul Healy, Philippe Jehiel, and James Peck. We thank John H. Kagel for generously providing the data for the Common Value Auction~\cite{avery1997second}.}
\authornote{Also with SciEcon CIC, London, United Kingdom, WC2H 9JQ}

\renewcommand{\shortauthors}{Dan Levin and Luyao Zhang}

\begin{abstract}
  We introduce NLK, a model that connects the Nash equilibrium (NE) and Level-K. It allows a player in a game to believe that her opponent may be either less or as sophisticated as, she is, a view supported in psychology. We apply NLK to data from five published papers on static, dynamic, and auction games. NLK provides different predictions than those of the NE and Level-K; moreover, a simple version of NLK explains the experimental data better in many cases, with the same or lower number of parameters. We discuss extensions to games with more than two players and heterogeneous beliefs.
  \\[10pt]
\end{abstract}


\maketitle

\section{Introduction}

There is mounting, robust evidence from laboratory experiments of substantial discrepancies between the prediction of Nash equilibrium (NE) and the behavior of agents.\footnote[1]{There is much experimental evidence that predictions of both (Bayesian) NE in static games and subgame perfect Nash equilibrium (SPNE) in dynamic games fail miserably. For instance, see~\citeN{mckelvey1992experimental}~and~\citeN{kagel2002information}}. Among all the alternative models that retain the individual rationality, but relax correct beliefs, Level-K is probably the most prominent model.\footnote[2]{Another strand of models such as quantal response equilibrium~\citeN{mckelvey1995quantal} retains correct beliefs but allow errors in the best response.}  First proposed by~\citeN{stahl1994experimental,stahl1995players}~and~\citeN{nagel1995unraveling}, Level-K introduces a non-equilibrium, structural model of strategic thinking, which admits possible cognitive limitations of players that are not allowed in NE.\footnote[3]{There are many variations and extensions of the Level-K model, and we refer the reader to~\cite{crawford2013structural} and the references therein.} This model has a hierarchy of levels of sophistication that are constructed iteratively, starting with an exogenous, nonstrategic, and least-sophisticated level$_0$ player. Higher levels are then constructed by assuming that a level$_{\mathrm{k}}$ player best responds to level$_{\mathrm{k}-1}$ opponents, $\text{k}=1,2, \ldots .$ Absent in NE, the Level-K model explicitly allows players to consider their opponents as less (strategically) sophisticated than themselves; however, it does not allow players to use a critical element of strategic thinking, namely, "put yourself in the other's shoes."
\par
Our paper introduces NLK, which bridges the NE and the Level-K model. Whereas a Nash player believes that the other player is another Nash player, and a Level-K player believes that the other player is less sophisticated than herself, NLK allows the player to believe, with a probability $\lambda$, the other player can be a naïve player, less sophisticated, than herself, and with a probability of (1-$\lambda$),  another NLK player, as sophisticated as herself. However, the NLK player still best responds to her subjective beliefs such as those in both NE and Level-K.\footnote[4]{The formal definition of NLK and its extensions to Bayesian and dynamic games are in Section 2.}   Next, we discuss how to construct a hierarchy of levels as Level-K does. 
\par
There are two possible interpretations of our model:	
\par
\begin{enumerate}
\item \textbf{A population game with inconsistent beliefs:\footnote[5]{ The population game interpretation provided in this section is simplified to the case where NLK model has only NLK players (i.e., NLK$_{1}$) facing Level$_{0}$ players. This is the simplest case that guides most of our data analysis.  To coincide with the second interpretations as a model of hierarchy of heterogeneous players the population game needs elaboration and is much less intuitive, so it is committed.}}  In this interpretation, an NLK player behaves as if facing a population composed of naïve and NLK agents. In equilibrium, an NLK player best responds to her belief that with a subjective probability $\lambda$, each of the other players in the game is a naïve player, and with a probability of (1-$\lambda$), each of the other players in the game is another NLK player (like herself). Being subjective, $\lambda$ does not have to coincide with the objective proportions of naïve players in the population, denoted by $\rho$. Thus, with $\lambda\neq\rho$, NLK is not a “full-equilibrium”\footnote[6]{~\citeN{stahl1995players}~include a rational expectation type along with different types of level-k and Nash players in analyzing experimental data of a 3×3 symmetric game. Their results reject the existence of a rational expectation type.} because it allows an NLK player to hold inconsistent beliefs regarding the proportion of naïve players in the population. As such, NLK belongs to the “bounded rationality” or behavioral models proposed to help reconcile the theoretical predictions and experimental and field evidence by maintaining the maximization and best response parts of individual rationality but allowing relaxation, in some form, of consistent beliefs, for example, CI 2007 and the other variations of the original Level-k model and the~\citeN{eyster2005cursed} model of \textit{cursed equilibrium}.
\par
Such inconsistency has been supported in psychology: The “False Consensus Effect,” first introduced by~\citeN{ross1977false}, claims that individuals overestimate the proportion of individuals similar to themselves ($\lambda<\rho$).\footnote[7]{ In the psychology literature, we found much support for the finding of FCE and the “self-anchoring” argument.~\citeN{mullen1985false} report on 115 studies that show FCE. For a more detailed empirical and theoretical discussion, refer to~\citeN{marks1987ten} and all the listed references therein.}   More recent works in psychology and experimental economics, have re-evaluated the “False Consensus Effect.” Some of those works have provided evidence in support of such effect~\cite{krueger1994truly,jimenez2019false}, but other works have demonstrated evidence of an opposite effect ($\lambda<\rho$;~\cite{dawes1990potential,sherman1984mechanisms})~or the absence of a biased belief (~\cite{engelmann2000false}). Of course, an individual may insist on consistency by requiring that in a “full-equilibrium,” $\lambda=\rho$.\footnote[8]{An alternative approach is to construct the naïve player's strategy based on the Poisson cognitive hierarchy $(\mathrm{P}-\mathrm{CH})$ model (~\cite{camerer2004cognitive}). ${ }^{9}$ We let $f(m)=\frac{e^{-\tau} \tau^{m}}{m !}$ denote the probability function of a Poisson distribution, and in a similar way as in the $\mathrm{P}-\mathrm{CH}$ model treatment of truncated probability distributions, we let an $\mathrm{NLK}_{\mathrm{m}}$ player believe that she faces a naïve player of level $h=0,1, \ldots, m-1$ with probability $g(h)=\frac{f(h)}{\sum_{l=0}^{m} f(l)}$ and another NLK $_{\mathrm{m}}$ player with probability $\left(1-\sum_{h=0}^{m-1} g(h)\right)=\frac{f(m)}{\sum_{l=0}^{m} f(l)}$.}
\par  
With $0<\lambda<1$, the sophistication of Level-K and NLK players are different and not easily ranked because NLK is a hybrid model “bridging” Level-K to the NE. The behavior of NLK is determined endogenously by having a player best respond to a mix of the naïve players, whose behavior is determined exogenously by the level$_0$ player in Level-K and of the NLK players. As an equilibrium model, it uses similar assumptions as in the NE, for example, all NLK players have mutual knowledge of rationality. In principle, the strategy of the naïve player can be specified as context based as long as it is exogenously given; and all levels of NLK are sophisticated compared with naïve players because their strategies are endogenously by (solving) by best responses.  
\item \textbf{A hierarchy of heterogeneous players (as an analog to the Level-K model):} A player is an NLK player of type $m$, denoted by NLK$_{m}$, who best responds to the belief that each of the other players is an NLK$_{m-1}$ player with probability $\lambda$ and another NLK$_{m}$ (like herself) with probability 1- $\lambda$. Thus, an NLK$_{m}$ player coincides with a levelm player when $\lambda$=1, and the NLK equilibrium reduces to NE when $\lambda$=0.
\end{enumerate}
To illustrate the NLK equilibrium, we consider a simple example of the chicken game introduced by~\citeN{rapoport1966game}. In this two-player symmetric game, each player chooses either a “Dove” or “Hawk,” and the player’s payoffs depend on her action and that of the opponents as follows: 
\par
\begin{table}[ht]
\resizebox{0.3\textwidth}{!}{%
\begin{tabular}{ccc}
     & Dove  & Hawk  \\ \cline{2-3}
Dove & \multicolumn{1}{|c|}{30,30} & \multicolumn{1}{c|}{20,70} \\
Hawk & \multicolumn{1}{|c|}{70,20} & \multicolumn{1}{c|}{0,0} \\ \cline{2-3}  
\end{tabular}%
}
\caption{Dove and Hawk Game.}
\label{tab:1}
\end{table}
\par
A random level$_{0}$ chooses to play either Dove or Hawk with equal probability. A level$_{1}$ best responds to the level$_{0}$ player by choosing Hawk, a level$_{2}$ best responds to the level$_{1}$ player by choosing Dove, a level$_{3}$ best responds to the level$_{2}$ player by choosing Hawk, and so forth. There are two pure NE strategies - (Hawk, Dove) and (Dove, Hawk) -and a third mixed strategy where Dove and Hawk are played with the probability of $(1/3)$ and $(2/3)$, respectively. Now, we consider an $\mathrm{NLK}_{1}$ player who faces a naïve random level$_{0}$ player with the probability of $\lambda$ and another $\mathrm{NLK}_{1}$ player with the probability of $(1-\lambda)$. For $\frac{2}{3} \leq \lambda \leq 1$, and only one pure strategy NLK equilibrium exists, where each player chooses Hawk. For $0 \leq \lambda<\frac{2}{3}$, there exist two pure strategy NLK equilibria-(Hawk, Dove) and (Dove, Hawk)–and a mixed strategy where Dove and Hawk are played with the probabilities of $\frac{2-3 \lambda}{6(1-\lambda)}$ and $\frac{4-3 \lambda}{6(1-\lambda)}$, respectively.

\textbf{NLK helps bridge Level-K to the NE both theoretically and empirically}:
\textbf{Theoretically}, it shares a similar foundation with NE but is also applicable to games with players of different cognitive or reasoning abilities. For example, in the experiment of~\citeN{alaoui2016endogenous}, math and science students who interact with students from the humanities field, may adopt a different subjective $\lambda$ than when they play with fellow math and science students. Such a conjecture, (e.g., larger $\lambda$) is reasonable and can be tested. We also adapt our basic definition of NLK to Bayesian games and dynamic games, as extensions of the Bayesian NE (BNE) and subgame perfect NE (SPNE). In the dynamic game, we show that NLK can characterize belief updating, which is absent in Level-K.
\par
\textbf{Empirically}, we compare the performance of NLK to that of NE and some versions of Level-K by applying it to data from four experimental papers published in top economic journals and to data from one field study. These studies allow us to test the NLK on a static game with complete information, a static game with incomplete information, and a dynamic game of perfect information and on field data. For the experiments that we analyzed, NLK provides several insightful implications. 
\par
First, in the static guessing game by~\citeN{arad201211}, a simple version of NLK with one parameter, $\lambda\in(0,1)$, that is chosen optimally, fits data better than both the NE and Level-K models with an optimal distribution among three types of players, namely, two parameters. When we allow for an error structure that is sensitive to payoffs but uses only one parameter, NLK still outperforms Level-K models. However, when we allow Level-K to freely choose more parameters, it fits better than the simple NLK, suggesting that in some cases, NLK can also be an analytical tool. 
\par
Second, when we apply the data from a centipede game experiment by~\citeN{palacios2009field}, NLK’s predictions, adapted to dynamic games, are different and more precise than those of the SPNE and Level-K models, with only a few exceptions when they coincide, or when Level-K adopts more parameters. Also reassuring that the optimal $\lambda<1$is the largest when the players are all students, the smallest when only chess players are involved, and in the middle when one chess player is matched with o student. Thus, the optimal $\lambda$ for NLK seems to track and capture the shift in subjective beliefs that can be expected in the different mixes of subject populations. The better performance of NLK than Level-K in the centipede game is reconfirmed by using the data from~\citeN{levitt2011checkmate}. Moreover, NLK can capture belief updating in every round of a game that a dynamic Level-K cannot. Notably, although the results of the data from the centipede game in the two aforementioned papers are drastically different, NLK predicts both quite well with different optimal $\lambda s$, which implies that the difference in behavioral data can be explained by the difference in beliefs of subjects between two datasets. 
\par
Third, we compare predictions of NLK to those of Level-K for the data from the common value auction experiment by~\citeN{avery1997second}. For inexperienced bidders, NLK’s performance coincides with that of Level-K; but for experienced bidders, NLK with $\lambda\in(0,1)$ provides the most accurate prediction. Moreover, because the estimated $\lambda$ is smaller for the data of experienced bidders compared with that of inexperienced bidders, NLK may also be used to track dynamic learning from experience, for example, learning in repeated games and convergence to a “full-equilibrium,” $\lambda=\rho=0$. 
\par
Forth, in recent experimental work on a rank-order tournament with an outside option - a dynamic game with imperfect information -~\citeN{brunner2020self} finds that a mixture of Level-K and NLK predicts both the population of types in the tournament and the mean-variance of efforts remarkably well. That paper shows that NLK predicts the experimental data better than a Level-K model without the updating of beliefs, which highlights the importance of the belief updating that PBNLK added to the Level-K and the validity of NLK for outside sample predictions. \footnote[9]{\citeN{brunner2020self} shows that the Nash equilibrium performs even worse than the Levelk without belief updating.}
\par
Finally, we also looked at data from the experimental work on the beauty contest game by~\citeN{gill2016cognitive} but concluded that there are no “winners” as the simplest Level-K and NLK models poorly predict behavior and Nash equilibrium performs much worse.

\textbf{Related Literature:}
Level-K and its related extension, the cognitive hierarchy models by~\citeN{camerer2004cognitive}, have been applied to many laboratory experiments and field data. The survey by~\citeN{crawford2013structural} documents many successes of Level-K and its extensions over other solution concepts, including NE. However, as we observed in the chicken game and in several examples in the following paper, NLK can be more useful than Level-K in certain games.
\par
Theoretically, Level-K has been extended in two ways.~\citeN{strzalecki2014depth} allows beliefs to vary arbitrarily for players at a certain level. Specifically, a level$_\text{k}$ player can believe the opponent to be level j $<$ k, by any arbitrary subjective distribution. However, and here as well, beliefs are restricted to lower levels. Building on~\citeN{strzalecki2014depth}, ~\citeN{jimenez2019false} innovatively allows a level$_\text{k}$ player to believe the opponent to be also a level$_\text{k}$ player, but only when their beliefs coincide, and adopts the solution concept of \textit{interim correlated rationality} to games of incomplete information that endogenize level-0 behavior. However, in the application to the \textit{e-mail game}, the case where the player allows the opponent to be at the same level is not considered. 
\par
\citeN{alaoui2016endogenous} use another approach and show how cognitive bounds, beliefs about opponents, and beliefs about opponents’ beliefs vary according to incentives by using a cost-benefit analysis. In their model, if agents believe that their opponents behave at lower levels than their own cognitive bound, they would behave at one level higher than these opponents; but if they believe that the strategies of their opponents are reaching or exceeding their own cognitive bound, they would act within their own cognitive bound. Thus, although the aforementioned researchers have considered a situation where the opponents have the same or even a higher cognitive level than the agents, they treated it as if the opponents were nevertheless one level below the agents. Thus, according to our review of the literature, no extension of the Level-K model allows either the player to believe she faces the same level as herself or apply such belief structure in the analysis of games. More recently,~\cite{koriyama2018inclusive} introduce a model of inclusive cognitive hierarchy that extends the cognitive hierarchy (CH) model in normal form games. Their model allows the most sophisticated player to consider the possibility that other players are of the same level as themselves. However, their model, unlike ours, requests that the belief of the most sophisticated player be consistent with objective distributions of player types. Our model subsumed their model in the extreme case where the strategy of the NLK naïve player is constructed as in CH and requires consistent belief. They explicitly recognize that such a requirement restricts the applicability of their model to most strategic environments such as beauty contests games, market-entry games, coordination games, and centipede games. They also do not characterize dynamic games and belief updates, which are covered in ours.
\par
NLK is not the first equilibrium solution concept to introduce an exogenous type; \citeN{kreps1982rational,kreps1982reputation,kreps1982sequential, milgrom1982theory, milgrom1982predation} (KMRW) have already used an exogenous type. However, NLK and KMRW’s models differ drastically in motivation and generality.
\par
\textbf{Motivation}: KMRW’s works have been motivated by~\citeN{selten1978} chain-store paradox (CSP) and by vast experimental evidence of cooperation in finitely repeated prisoner’s dilemma (PD) games. \textit{Deterrence strategy} in CSP  and \textit{cooperation} in PD games contradict the logic of backward induction that implies unraveling to the one-shot, stage game, solution. KMRW’s objective is to resolve the paradoxes of using a deterrence strategy in the CSP\footnote[10]{\textit{Deterrence strategy}, where the monopoly fights an early entrant, although it is not the best response in the stage game, was offered by~\citeN{selten1978}, as a sensible, though not an equilibrium, strategy to deter later entrant.} game and cooperation in the finitely repeated PD game. To do so, they transform these complete games, into incomplete, information games by introducing a “tiny” probability of exogenous type and showing that it is sufficient to “choke off” the otherwise unavoidable logic of unraveling. The emphasis on tiny probability is a critical novelty, because otherwise, the deterrence strategy in the CSP game or cooperation in the PD game may be rationalized even in a one-shot game. In NLK, the probability $\lambda$ of such an exogenous type is typically quite large and similar to, but maybe smaller than, that in Level-K model. Thus, whereas the motivation of KMRW’s models is to “defend” the standard NE, NLK is a behavioral model of bounded rationality. 
\par
\textbf{Generality}: NLK introduces one nonstrategic exogenous type to be applied to all, or at least to a large class of different, games. By contrast, KMRW admit that their “defense” of the standard NE requires a particular exogenous type for each case.\footnote[11]{KMRW explicitly acknowledge that a particular, and different, exogenous type may be necessary for different cases.}  For instance, in the CSP case, Kreps and Wilson use a “strong” monopoly that is hard-wired to fight; in the finitely repeated PD game, KMRW use two nonstrategic types for two cases, respectively: the one who plays Tit-for-Tat in the one-sided incomplete information game, and the one who prefers the stage payoffs from joint cooperation over those of defection when the other player cooperates, for their two-sided case.\footnote[12]{In addition, NLK can require that $\lambda$ matches the probability of the exogenous type in the population, making the model an equilibrium model with rational expectations.} 
\par
Similar to other models that use “relaxed beliefs,” NLK has its limitations. For example, NLK cannot explain deviations from theoretical predictions in games with a dominant strategy solution, such as overbidding the value in second-price sealed-bid auctions with private values, first reported by~\citeN{kagel1987}.
\par	
In Section 2, we present our basic solution concepts as used in different types of games (static or dynamic, with complete or incomplete information). In Sections 3, 4, and 5, we provide the NLK solutions and compare them to those of the NE and Level-K models for a static guessing game, a dynamic centipede game, and a common value auction. We conclude in Section 6. Readers can refer to Appendix A.3 for an extended discussion for solution concepts related to NLK.  
\section{The solution concept}
In this section, we define the NLK equilibrium in a simple case with symmetric beliefs for three types of games and prove existence. We discuss several extensions in Section 5.
\subsection{Basic Case}
We consider that an N-player normal form game $G=\left(i, S_{i}, u_{i}\right)_{i \in I}$ comprises a set $I=\{1,2, \ldots, N\}$ players, where $\left(S_{i}, u_{i}\right)$ are the strategy set and the utility function of player $i$, respectively. The strategy of a naïve player $i$ is given exogenously by $\sigma_{i}^{o} \in \Delta\left(S_{i}\right), i \in I .$ In our strategic environment, an NLK player believes that each of the opponents is either a naïve player with probability $\lambda$, or another NLK player with probability $(1-\lambda), \lambda \in[0,1] .$ In equilibrium, an NLK player chooses an optimal strategy by best responding to her belief. A formal definition of the NLK equilibrium is as follows:

\vspace{6pt}

\noindent \textbf{Definition 1.} \textit{A mixed strategy profile} $\left(\sigma_{i}^{*}\right)_{i \in I}$, \textit{is a} $\lambda-N L K$ \textit{equilibrium if for each} $i \in I$, \textit{and each} $s_{i}^{\prime} \in S_{i}, \lambda u_{i}\left(\sigma_{i}^{*}, \sigma_{-i}^{0}\right)+(1-\lambda) u_{i}\left(\sigma_{i}^{*}, \sigma_{-i}^{*}\right) \geq \lambda u_{i}\left(s_{i}^{\prime}, \sigma_{-i}^{0}\right)+(1-\lambda) u_{i}\left(s_{i}^{\prime}, \sigma_{-i}^{*}\right).$\footnote[13]{This version has homogenous NLK players; allowing heterogonous players may help better fit the model to the data it also introduces additional parameter(s) and reduces the transparency that we wish to maintain.}

\subsection{Bayesian Games}

We consider a Bayesian game of incomplete information, $B=\left(S_{i}, u_{i}, \Theta_{i}, p\right)_{i \in I}$, where $\Theta_{i}$ denotes the set of player $i$ 's types and $p$ denotes the joint density function of the probability distribution over $\prod_{i \in I} \Theta_{i}$. Similar to the relationship between NE and BNE, a BNLK is the NLK equilibrium of the "extended game" in which each player $i$ 's space of pure strategies, $S_{i}{ }^{\Theta_{i}}$, denotes the set of mappings from $\Theta_{i}$ to $S_{i}$. Again, we let $\sigma_{i}^{0} \in \Delta\left(S_{i}\right), i \in I$, denote the strategy of a naïve player $i$, which is independent of that player type. A formal definition of BNLK with symmetric subjective belief $\lambda$ is as follows.

\vspace{6pt}

\noindent\textbf{Definition 2.} \textit{A profile of strategies} $\left\{s_{i}^{*}(\cdot)\right\}_{i \in I}$, \textit{is a} $\lambda-B N L K$ \textit{equilibrium, if for each} $i, i \in I$, \textit{and each} $\theta_{i} \in \Theta_{i},$ 
\newline

$s_{i}^{*}\left(\theta_{i}\right) \in \arg \max _{s_{i} \in S_{i}} \int p\left(\theta_{-i} \mid \theta_{i}\right)\left[\lambda u_{i}\left(s_{i}, \sigma_{i}^{0} ; \theta_{i},\right.\right.
\theta_{-i}) \left.+(1-\lambda) u_{i}\left(s_{i}, s_{-i}^{*}\left(\theta_{-i}\right) ; \theta_{i}, \theta_{-i}\right)\right] d \theta_{-i}
$

\subsection{Dynamic Games}

Consider a dynamic game with perfect information and perfect recall,\footnote[14]{That is, at any decision node, all previous moves are assumed to be known to every player.} $ P=\left(u_{i}, \Upsilon\right)_{i \in I}$,
where $\Upsilon$ denotes a game tree and $I$ is the set of players. A node in $\Upsilon$ is denoted by $h^{t}$, and the set of nodes is denoted by $H$. The set of nodes at which player $i$ must move is denoted by $H_{i}$. An
NLK player holds a prior belief that the each of the other players is either a naïve player with probability $\lambda$ or another NLK player with probability $(1-\lambda), \lambda \in[0,1] .$ At every decision node with history $h^{t}$, as more information is revealed, beliefs are updated. We denote the updated belief that the opponent is a naïve player by $p_{i}\left(h^{t}\right) .$ In equilibrium, an NLK player chooses an optimal strategy according to her belief at every decision node and the choice is sequentially rational as in Definition 3:

\vspace{6pt}

\noindent \textbf{Definition 3.} (Sequential rationality). \textit{A strategy profile} $\left\{\sigma_{i}^{*}\right\}_{i \in I}$ \textit{is sequentially rational with respect to the profile of beliefs} $\left\{p_{i}\left(h_{i}^{t}\right)\right\}_{h_{i}^{t} \in H_{i}}, i \in I$ \textit{if for} $i \in I$ \textit{all strategies} $\sigma_{i}^{\prime}$, \textit{and all nodes} $h_{i}^{t} \in H_{i}$:

\begin{equation}
\begin{aligned}
&p_{i}\left(h_{i}^{t}\right) u_{i}\left(\sigma_{i}^{*}, \sigma_{-i}^{0} \mid h_{i}^{t}\right)+\left(1-p_{i}\left(h_{i}^{t}\right)\right) u_{i}\left(\sigma_{i}^{*}, \sigma_{-i}^{*} \mid h_{i}^{t}\right) \geq p_{i}\left(h_{i}^{t}\right) u_{i}\left(\sigma_{i}^{\prime}, \right.\\
&\sigma_{-i}^{0} \mid h_{i}^{t}) + \left(1-p_{i}\left(h_{i}^{t}\right)\right) u_{i}\left(\sigma_{i}^{\prime}, \sigma_{-i}^{*} \mid h_{i}^{t}\right)
\end{aligned}
\end{equation}

We also require that the beliefs of NLK players be consistent. That is, they may start with a subjective prior distribution and then be updated by the Bayes' rule at each succeeding decision node. To present formally the consistency restriction, we let $p\left(h^{t} \mid \sigma_{i}, \sigma_{-i}\right)$ denote the probability that decision node $h_{t}$ is reached according to the strategy profile, $\left(\sigma_{i}, \sigma_{-i}\right)$.

\vspace{6pt}

\noindent \textbf{Definition 4.} (Consistency). \textit{A profile of beliefs} $\left\{p_{i}^{*}\left(h_{i}^{t}\right)\right\}_{h_{i}^{t} \in H_{i}} i \in I$ \textit{is consistent with the subjective prior} $\lambda$ \textit{and the strategy profile} $\left\{\sigma_{i}\right\}_{i=1,2}$ \textit{if and only if for} $i \in I$, \textit{and all nodes} $h_{i}^{t} \in H_{i}$:

\begin{equation}
p_{i}^{*}\left(h_{i}^{t}\right)=\frac{\lambda p\left(h_{i}^{t} \mid \sigma_{i}, \sigma_{-i}^{0}\right)}{\lambda p\left(h_{i}^{t} \mid \sigma_{i}, \sigma_{-i}^{0}\right)+(1-\lambda) p\left(h_{i}^{t} \mid \sigma_{i}, \sigma_{-i}\right)},
\end{equation}

Where, $p\left(h_{i}^{t} \mid \sigma_{i}, \sigma_{-i}^{0}\right)>0$ or $p\left(h_{i}^{t} \mid \sigma_{i}, \sigma_{-i}\right)>0.16$ \footnote[15]{Notably, Definition 1.4 places no restrictions on player i’s expectations on those decision nodes that cannot be reached according to $\sigma$, regardless of whether the player faces a naïve player or another NLK player. A stronger notion of consistency could be defined in the spirit of a trembling hand or a sequential equilibrium~\citeN{kreps1982sequential}. Such a stronger restriction and its impact on prediction are discussed in Section 5.}

Although the game has perfect information, the belief structure in our strategic environment makes our solution concept more similar to an analogy of a perfect Bayesian equilibrium; thus, we denote it as PBNLK and formally treat it as follows:

\vspace{6pt}

\noindent \textbf{Definition 5.} An assessment $\displaystyle\left(\sigma_{i}^{*},\left\{p_{i}^{*}\left(h_{i}^{t}\right)\right\}_{h_{i}^{t} \in H_{i}}\right)_{i \in I}$ is a $\lambda-PBNLK$ equilibrium if

\begin{enumerate}
    \item the strategy profile $\left\{\sigma_{i}^{*}\right\}_{i \in I}$ is sequentially rational with respect to the profile of beliefs $\left\{p_{i}^{*}\left(h_{i}^{t}\right)\right\}_{h_{i}^{t} \in H_{i}}, i \in I$, and
\item the profile of beliefs $\left\{p_{i}^{*}\left(h_{i}^{t}\right)\right\}_{h_{i}^{t} \in H_{i}}, i \in I$ is consistent with the subjective prior $\lambda$ and the strategy profile $\left\{\sigma_{i}^{*}\right\}_{i \in I}$
\end{enumerate}

\subsection{Existence}

\noindent \textbf{Proposition 1.} for any $\lambda\in[0,1]$:
\begin{enumerate}[label=\alph*)]
    \item In every finite strategic-form game, there exists an NLK equilibrium.
\item In every finite Bayesian game, there exists a BNLK equilibrium.
\item In every finite extensive form game, there exists a PBNLK equilibrium.
\end{enumerate}
\vspace{6pt}
\noindent \textbf{Proof:} See Appendix A.1.
\par
In the following sessions, we compare the performance of NLK. We employ only k=1 to that NE or Level-K model with k$\geq$1,  where the naïve player is a random level$_0$ player that chooses uniformly among her strategy set across all games. 
\section{Arad-Rubinstein Money Request Game.}
In the basic version of the money request game by~\citeN{arad201211}, there are two risk-neutral players, and each can request and receive an integer amount of money from \$11 to \$20, plus an extra \$20 if she asks for exactly one integer less than the other player. 
\par
\begin{table}[ht]
\renewcommand{\arraystretch}{1.5}
\begin{tabular}{cccccccc}
\hline NLK (\%) $(\lambda)$ & 15 & 16 & 17 & 18 & 19 & 20 \\
\hline$\left[0 \leq \lambda \leq \frac{1}{2}\right)$ & $\frac{5(5-10 \lambda)}{1-\lambda}$ & $\frac{5(5-2 \lambda)}{1-\lambda}$ & $\frac{5(4-2 \lambda)}{1-\lambda}$ & $\frac{5(3-2 \lambda)}{1-\lambda}$ & $\frac{5(2-2 \lambda)}{1-\lambda}$ & $\frac{5(1-2 \lambda)}{1-\lambda}$   \\
{$\left[\frac{1}{2} \leq \lambda \leq \frac{14}{20}\right)$} & 0 & $\frac{5(14-20 \lambda)}{1-\lambda}$ & $\frac{15}{1-\lambda}$ & $\frac{10}{1-\lambda}$ & $\frac{5}{1-\lambda}$ & 0 \\
{$\left[\frac{14}{20} \leq \lambda \leq \frac{17}{20}\right)$} & 0 & 0 & $\frac{5(17-20 \lambda)}{1-\lambda}$ & $\frac{10}{1-\lambda}$ & $\frac{5}{1-\lambda}$ & 0 \\
{$\left[\frac{17}{20} \leq \lambda \leq \frac{19}{20}\right)$} & 0 & 0 & 0 & $\frac{5(19-20 \lambda)}{1-\lambda}$ & $\frac{5}{1-\lambda}$ & 0 \\
{$\left[\frac{19}{20} \leq \lambda \leq 1\right]$} & 0 & 0 & 0 & 0 & 100 & 0 \\
\hline
\end{tabular}
    \caption{NLK equilibrium strategy for different subjective beliefs.}
    \label{tab:2}
\end{table}
\par
Consider the Level-K model with a level0 payer who randomizes uniformly within the strategy set: {\$11,\$12,...,\$20}. A level$_{1}$ player that requests \$20 earns \$20. Alternatively, if she asks for \$19, she earns \$19 for sure and a \$20 bonus with a probability of 1/10, for a total expected payoff of \$21.\footnote[16]{To ask for any amount of money less than \$19 leads to a strictly lower payoff.} Thus, level$_{1}$ picks \$19, level$_{2}$ picks \$18,..., and level$_9$ picks \$11; but then, level$_{10}$ picks \$20, level$_{11}$ picks \$19, and so forth. Thus, it is difficult to infer from players’ actions their sophistication level: A player who requests \$19 can be a level$_{1}$ player or a highly sophisticated level$_{11}$ player. 
\par
Table~\ref{tab:2} shows the unique mixed strategy $\lambda$-NLK equilibrium for each $\lambda\in[0,19/20)$ and the unique pure strategy for $\lambda\in[19/20,1]$.\footnote[17]{See Appendix A.2 for detail of the argument.}   Table~\ref{tab:3} compares the performance of the Level-K model; k=1,2,3; NE; and NLK\footnote[18]{Our level$_0$, player is defined as a player who pick each available action with equal probability.}  by using the mean squared error (MSE). NLK with the best $\lambda$=0.6585, fits the data better than NE and any type of the Level-K model; moreover, it also outperforms Level-K with the optimal distribution of levels 1, 2 and 3, that is, two parameters, because it reduces MSE by 23.457\%. (from MSE being 35.93 to 28.39.)  We include the prediction of NLK where $\lambda$ is restricted to match $\rho$, but allowing $\lambda=\rho=0.70$, that minimizes

\begin{table}[ht]
\renewcommand{\arraystretch}{1.2}
\resizebox{0.8\textwidth}{!}{%
\begin{tabular}{cccccccccccc}
\hline
Action                                                                           & 11  & 12  & 13  & 14  & 15  & 16   & 17   & 18   & 19   & 20  & MSE   \\
level$_{1}$ (\%)                                                                      & 0   & 0   & 0   & 0   & 0   & 0    & 0    & 0    & 100  & 0   & 980.2 \\
level$_{2}$ (\%)                                                                      & 0   & 0   & 0   & 0   & 0   & 0    & 0    & 100  & 0    & 0   & 620.2 \\
level$_{3}$ (\%)                                                                      & 0   & 0   & 0   & 0   & 0   & 0    & 100  & 0    & 0    & 0   & 580.2 \\
\begin{tabular}[c]{@{}c@{}}\textit{level}$_k$,\textit{k}=1,2,3,\\ \textit{optimal distribution}\end{tabular} & 0   & 0   & 0   & 0   & 0   & 0    & 40.7 & 38.7 & 20.6 & 0   & 35.93 \\
NE (\%)                                                                          & 0   & 0   & 0   & 0   & 25  & 25   & 20   & 15   & 10   & 5   & 137.2 \\
\begin{tabular}[c]{@{}c@{}}NLK (\%)\\ $\lambda$=0.6585\end{tabular}                      & 0   & 0   & 0   & 0   & 0   & 12.1 & 43.9 & 29.4 & 14.6 & 0   & 28.39 \\ \begin{tabular}[c]{@{}c@{}} $level_0$  and NLK (\%)\\ ($\lambda=\rho=0.70$)\end{tabular}
 & 7   & 7   & 7   & 7   & 7   & 7    & 22   & 17   & 12   & 7   & 38.20 \\
\begin{tabular}[c]{@{}c@{}} $level_0$  and NLK (\%)\\ ($\rho$=0.344,$\lambda$=0.734)\end{tabular}                            & 3.4 & 3.4 & 3.4 & 3.4 & 3.4 & 3.4  & 32.  & 28.1 & 15.7 & 3.4 & 5.58  \\
Data (\%)                                                            & 4   & 0   & 3   & 6   & 1   & 6    & 32   & 30   & 12   & 6   &    \\ \hline  
\end{tabular}%
}
\caption{11-20 Game: Comparison of different solution concepts by MSE.}
\label{tab:3}
\end{table}
\par
the MSE.\footnote[19]{Here is how the entry of 22 in this row is calculated: For $\lambda=0.7$, we obtain from Table~\ref{tab:2}, that an NLK player pick the $\# 17, \frac{5(17-20 \lambda)}{1-\lambda}=50 . \rho=0.7$ implies that $30 \%$ are NLK players, resulting in $15(\%)$ coming from NLK players. We add $7(\%)$ coming from the $70 \%$ of Level-0 players who uniformly randomize over the ten numbers.}  As expected, it increases the unrestricted MSE, from 28.39 to 38.20, which is higher than 35.93 (by about 6\%) obtained  by the Level-K model with $k=1,2,3$ , and picking the optimal distribution of those levels, i.e., using 3 parameters. In contrast, the Level-K model that optimal distribution of those levels, i.e., using 3 parameters. In contrast, the Level-K model that uses only one level, $K=1$,or $K=2$ or $K=3$, results in MSE that is between 15 to 20 times higher.  We also include prediction of NLK using MSE minimizer pair of $\lambda$=0.734 and $\rho$=0.344 and this results in MSE of 5.58, which one fifth of the aforementioned MSE of Level-K.\footnote[20]{Observing that the data from~\citeN{arad201211} in Table~\ref{tab:3}, a referee noted, that $30 \%$ to $40 \%$ of the players "look like" Level-0 players and suggested having those prediction. Our "best" fitting $\rho=0.344$ is in the middle of that range.}  
\par
Finally, we test the robustness of our results by using an alternative statistical method. Our econometric specification follows the mixture-of-types models of~\citeN{stahl1994experimental,stahl1995players}.\footnote[21]{The same econometric specification was also adopted by~\citeN{costa2001cognition,camerer2004cognitive,costa2006,crawford2007level}. The error model is developed from quantal response equilibrium (See~\citeN{goeree2008regular} and discussed in~\citeN{goeree2001ten}.} 
\par
Both level$_\text{k}$ and our NLK types are assumed to make logistic errors as follows. The decision rule suggests that the choice probabilities of type t players are positive but imperfect and related to expected payoffs according to the specific beliefs of type t. Formally, we denote the expected payoff player i of type t, given strategy s by $\pi_i^t(s)$. Then, the probability of observing s by such players is specified as follows:
\par
\begin{equation*}
p_{i}^{t}(s)=\frac{\exp \left(\eta \pi_{i}^{t}(s)\right)}{\sum_{s^{\prime} \in s_{i}} \exp \left(\eta \pi_{i}^{t}\left(s^{\prime}\right)\right)}
\end{equation*}
\par
where $S_{i}$ and $\eta$ are respectfully the strategy set for player $i$, and the precision parameter.
Specifically, $\eta$ determines the sensitivity of the choice probabilities to payoff differences.\footnote[22]{As $\eta$ goes to $\infty$, the probability of the optimal decision converges to 1, i.e., the choice is error-free and fully characterized by the model under consideration. In contrast, as $\eta$ goes to 0, the choice probability converges to a uniformly random choice, such as that of the random level$_0$ players.}

Exceptionally, random level$_{0}$ directly specifies a uniform distribution of decisions and thus has no precision parameter. Alternatively, it is equivalent by specifying the precision parameter to be 0 for a random level$_{0}$ player. The likelihood of observing a sample $\left\{s_{i}\right\}_{i=1}^{N}$, given type $t$, is $L^{t}\left(\left\{s_{i}\right\} \mid \eta\right)=\prod_{i=1}^{N} p_{i}^{t}\left(s_{i}\right)$

Let $\alpha_{t}$ denote the proportion of type $t$ in the population, with $\sum_{t} \alpha_{t}=1$. The likelihood of observing the sample unconditional on type is $\prod_{i=1}^{N} \sum_{t} \alpha_{t} p_{i}^{t}\left(s_{i}\right) .$ Table~\ref{tab:4} reports the results. With

\begin{table}[ht]
\resizebox{0.8\textwidth}{!}{%
\begin{tabular}{ccccc}
\hline
Action & Log-Likelihood   (LL) & Precision parameter ($\eta$)                                 & BIC\footnote[23]{$BIC=kln(n)-2LL; k$ is the number of free parameters to be chosen, and n is the number of observations. }     & AIC\footnote[24]{$AIC=2k-2LL; k$ is the number of free parameters to be chosen.}     \\
level$_{1}$ & -233.970              & \begin{tabular}[c]{@{}c@{}}0.296\\ (0.039)\end{tabular} & 472.622 & 469.940 \\
level$_{2}$ & -226.245              & \begin{tabular}[c]{@{}c@{}}0.066\\ (0.009)\end{tabular} & 457.172 & 454.490 \\
level$_{3}$ & -221.220              & \begin{tabular}[c]{@{}c@{}}0.075\\ (0.010)\end{tabular} & 447.122 & 444.440 \\
\begin{tabular}[c]{@{}c@{}} level\_k,k=1,2\\ optimal distribution\end{tabular}       & -218.100              & \begin{tabular}[c]{@{}c@{}}0.252\\ (0.052)\end{tabular} & 445.564 & 440.200 \\
\begin{tabular}[c]{@{}c@{}}level\_k,k=1,2,3,\\ optimal distribution\end{tabular}       & -197.770              & \begin{tabular}[c]{@{}c@{}}0.207\\ (0.051)\end{tabular} & 409.586 & 401.540 \\
NE     & -230.040              & \begin{tabular}[c]{@{}c@{}}0.231\\ (0.046)\end{tabular} & 464.762 & 462.080 \\
\begin{tabular}[c]{@{}c@{}} NLK \\ ($\lambda$=0.85)\end{tabular}       & -210.050              & \begin{tabular}[c]{@{}c@{}}0.359\\ (0.025)\end{tabular} & 429.464 & 424.100 \\ \hline
\end{tabular}%
}
\caption{Comparison of different solution concepts by maximum log-likelihood }
\label{tab:4}
\end{table}
\par
An error structure, the best single type Level-K model, with $k^{*}=3$ has a smaller log-likelihood and a precision parameter, $L L=-221.275, \eta=0.075$ than those of NLK with the best $\lambda^{*}=$ 0.85: $L L=-210.05, \eta=0.359 .$ NLK also outperforms Level-K model with the optimal distribution of level$_{1}$ an'd level$_{2}$ of $L L=-218.093, \eta=0.252.$\footnote[25]{It is estimated to be the 85\% level$_1$ and 15\% level$_2$ types. } However, we let Level-K use two parameters, and optimal distribution, of level$_{1}$ level$_{2}$ and level$_{3}$, and this raises its $\mathrm{LL}$ to $-197.77$, which is larger than that of NLK with only one parameter, $-210.05.$ However, NLK still has higher precision, $\eta=0.231$, than that of Level-K, $\eta=0.207.$\footnote[26]{It is estimated to be the 46\% level1, 24.45\% level2, and 28.98\% level$_3$ types. } The results are robust when considering the Bayesian information criterion (BIC) and AIC instead of LL.

\section{Centipede game}
Introduced by~\cite{rosenthal1981games}, the centipede game is an example where deviations from \textit{backward induction} (or SPNE) seem reasonable.\footnote[27]{For additional literature, see~\citeN{mckelvey1992experimental, fey1996experimental, nagel1998experimental, bornstein2004individual,RAPOPORT2003239}. These papers show that even in high-stakes situations, involving altruism or group decisions, \textit{Backward Induction} remains inadequate to explain players’ behavior.}  We follow the bulk of the literature and study a version of the centipede game where the total payout doubles when the game continues to the next stage, which subsumes the game in the experiments of~\citeN{palacios2009field} and~\citeN{levitt2011checkmate}, as a special case (with six decision nodes).
\par	
There are two players, A and B, with an initial pot worth \$5. At Node 1, Player A moves and chooses either to stop the game (T) by taking 80\% of the pot and leaving 20\% of it to Player B or passes the game (P) to Player B, doubling the pot. If Player A chooses P, at Node 2, Player B faces a similar decision but with a pot now worth \$10. Unless one of the players chooses T earlier, the game ends after S=2N stages, with Player B either choosing T, taking 80\% of the pot and leaving the other 20\% to Player A, or choosing P and doubling the pot, with the result that 20\% of the pot goes to Player B and 80\% of it goes to Player A. The payoffs for Players A and B are  $(\$2^{2k},\$2^{sk-2}$ if the game ends at an odd decision node, 2\textit{k}-1, and $\$2^{2k-1},\$2^{sk+1}$ if the game ends at an even decision node, $2k, k=1,2,...,N-1$. By backward induction, the unique SPNE strategy profile for Player A is to play T immediately, at Node 1, and off equilibrium, the active player always chooses T at each Node.
\par	
Based on the dynamic Level-K model by~\citeN{ho2013dynamic}, it is equally likely that a level0 player chooses T or P at each decision node, and strategies of $K>0$ are generated from iterative best responses to a player of one level below. A level$_1$ Player B would choose T at the last node.\footnote[28]{To end the game at Node 2N, Player B receives a payoff of $\$2^{2N+1}$, and he only ends up with $\$2^{2N}$ if he chooses P instead.}
\par
We denote the whole pie at each decision node by $x.$ A level$_{1}$ Player A is playing T at node $(2N-1)$ yields $\frac{4x}{5}$, and playing $\mathrm{P}$ yields $\frac{9x}{5}$; thus, a level$_{1}$ Player A would choose $\mathrm{P}$ at the decision node
$(2N-1) $ Similarly, a level$_{2}$ Player A would choose $T$ at the penultimate node $(2 N-1)$
\par
Table~\ref{tab:5} summarizes the solution for the Level-K model for a game of length $S=2 N$. For a certain level of players (indicated in the second column), there exists a corresponding threshold stage (indicated in the first column). A level$_{\mathrm{k}}$ player chooses $\mathrm{P}$ before the threshold stage $s^{*}$ but chooses T at stage $s^{*}$ and afterward. For example, in a six-stage game $(N=3)$, the threshold stage for a level$_{3}(k=3, h=1)$ Player A is $2(3-1)+1=5 .$ Thus, a level$_{3}$ Player A chooses $\mathrm{P}$ before Node 5 and T at Node 5 .
\par
\begin{table}[ht]
\centering
\resizebox{0.6\textwidth}{!}{%
\begin{tabular}{cccccll}
\cline{3-5}                  &                   & Role                     & Threshold stage s                 & Level of players                 &                  &                  \\ \cline{3-5}
                  &                   & Player A                 & 2(N-h)+1                          & \begin{tabular}[c]{@{}c@{}} $k=2h^*$  \textit{or} $2h+1$ \\ $(1\leq h\leq N-1)$ \end{tabular}                                  &                  &                  \\ \cline{3-5}
                  &                   & Player B                 & 2(N-h)+2                          &        \begin{tabular}[c]{@{}c@{}} $k=2h^*-1$  \textit{or} $2h$ \\ $(1\leq h\leq N)$ \end{tabular}                                      &                   &                  \\ \cline{3-5}
\multicolumn{7}{l}{\begin{tabular}[c]{@{}l@{}}$h^*$ is an auxiliary parameter for indicating the same threshold \\stage of two adjacent levels.\end{tabular}}
\end{tabular}%
}
\caption{Threshold stage for different levels of players.}
\label{tab:5}
\end{table}
\par
In general, a Player A, at level $k=2N$ or higher, and a Player B at $k=(2N-1)$ or higher, ought to choose T at each decision node. The Level-K solution requires relatively high levels\footnote[29]{Table~\ref{tab:5} also entails that to increase the level by just 1 would not necessarily predict earlier termination. Two adjacent levels of players might behave in the same manner. }  to rationalize terminating the game at earlier stages, especially for longer games, because the strategies of different level players are independent of the length of the game. For example, regardless of the duration of the game, a level1 Player A ought to keep passing to the last decision node, and regardless of the observed history, a level$_\text{k}$ player never updates his belief.\footnote[30]{Note that in a more general CH solution concept may produce qualitatively different predictions. However, because beliefs put more weight on lower levels according to a Poisson distribution in CH and lower levels continue passing to later stages, an even higher level of players than in the Level-K model would be required to rationalize early termination.} 

Consider a simple version of PBNLK with symmetric beliefs, $0<\lambda<1$. At the last stage, T is the best response for Player B regardless of his belief about his opponents’ type. We now assume that Player B first chooses T at Stage 2$n$ and Player A plans to choose T at Stage ($2n+1$). Then, at stage ($2n-1$), Player A’s posterior belief of the opponent being level$_0$ is 

\begin{equation}
p_{A}^{\lambda}(2 n-1)=\frac{\lambda\left(\frac{1}{2}\right)^{n-1}}{\lambda\left(\frac{1}{2}\right)^{n-1}+(1-\lambda)}=\left[\frac{\left(\frac{1}{2}\right)^{n-1}}{\left(\frac{1}{2}\right)^{n-1}+\frac{(1-\lambda)}{\lambda}}\right] \in(0, \lambda)
\end{equation}

If Player B first plays T at Stage $2n$, at Stage ($2n-1$), Player A receives 4x/5 by playing T, whereas by playing P now and T at  ($2n+1$) yields the expected payoff:

\begin{equation*}
\left[\frac{p_{A}^{\lambda}(2 n-1)}{2}+1-p_{A}^{\lambda}(2 n-1)\right] \frac{2 x}{5}+p_{A}^{\lambda}(2 n-1) \frac{1}{2} \times \frac{4}{5} \times 4 x=\frac{2 x}{5}+p_{A}^{\lambda}(2 n-1) \frac{7 x}{5}
\end{equation*}

Thus, Player A plays $\mathrm{P}$ whenever $\frac{2}{7}<p_{A}^{\lambda}(2 n-1) \leq 1$ and plays $\mathrm{T}$ otherwise. Moreover, because $p_{A}^{\lambda}(2 N-1)$ decreases in $N$ for a given $\lambda$, in a longer game, NLK Player A (with a certain $\lambda)$ is more likely to play T at stage $(2 N-1)$. This result is a key difference between NLK and the Level-K model where a level$_{1}$ Player A always passes at stage $(2 N-1)$ regardless of the game's duration. Because $p_{A}^{\lambda}(2 n-1)(\leq \lambda)$ is strictly decreasing in $n$, and $p_{A}^{\lambda}(2 n-1)_{n \rightarrow \infty}=0$, for $\lambda \leq \frac{2}{7}$, Player A would always play $\mathrm{T}$, given that Player $\mathrm{B}$ plays $\mathrm{T}$ in the next sage. For $\lambda>\frac{2}{7}$, by continuity, there is a critical value $n_{A}$, such that $p_{A}^{\lambda}(2 n-1)>\frac{2}{7}$ for $n<n_{A}$, and $p_{A}^{\lambda}(2 n-1) \leq \frac{2}{7}$ for $n \geq n_{A} .$

Similarly, we assume that Player A first chooses T at Stage $(2 n+1),(n \leq N-1)$ and Player B plans to choose $\mathrm{T}$ at stage $(2 n+2)$. Then, at Stage $2 n$, Player B's posterior belief that the opponent is level$_0$ is

\begin{equation*}
p_{B}^{\lambda}(2 n)=\frac{\lambda\left(\frac{1}{2}\right)^{n-1}}{\lambda\left(\frac{1}{2}\right)^{n-1}+(1-\lambda)}=p_{A}^{\lambda}(2 n+1)
\end{equation*}

This implies that the threshold stage for Player B, $s_{B}^{*}$, is one stage earlier than that of Player A, $s_{A}^{*}$, that is, $s_{B}^{*}=\left(2 n_{B}\right)=\left(2 n_{A}-1\right)-1=s_{A}^{*}-1$

We use these arguments to construct our PBNLK equilibrium. For $\lambda=0$, the game ends at the first stage (the same result as in SPNE).\footnote[31]{The off-equilibrium path will not be reached by an A or B NLK player whether her opponent is an NLK or a naive player; thus, it is not restricted by Definition 1.4 of consistency. We assume that an NLK player believes the other NLK player would always play T off the equilibrium path.} For $\lambda>0$, there are two possibilities. In a short game with a relatively larger $\lambda$ satisfying $p_{A}^{\lambda}(2 N-1)>\frac{2}{7}$, Player A plays $\mathrm{P}$ to the end, and Player B first plays T at the last stage (the same result as when both players are level$_{1}$ ). In a longer game with $p_{A}^{\lambda}(2 N-1) \leq \frac{2}{7}$, the game would end earlier. For similar arguments as in the papers of KMRW (1982),\footnote[32]{Inserting a “crazy” type even with a slight probability can rationalize long cooperation in the finitely repeated prisoners’ dilemma games. } PBNLK must be in mixed strategies for this range of $\lambda$. The reason for this is that in a presumed pure strategy, PBNLK and an NLK player (who ought to play T earlier than the other player) would rather deviate in the first node, that is, she ought to play T and lay P instead. Completing this action would mislead the other player to believe that he is facing a levelo player (as only a levelo player would have played $\mathrm{P}$ in the last node); thus, the other NLK player would play P.\footnote[33]{For example, in the case when the threshold stage of Player B is 4 and (it follows) that of Player A is 5, now, at Node 5, which is reachable for Player A when facing a level0 player, since Player B first choose T at 4, not 6, the belief $P_A^\lambda$ (5) represented by Equation 1.3 no longer satisfies our consistency requirement. Upon reaching Node 5, by Bayes’ rule, an NLK Player A confirms that her opponent is a level0 player for sure, so she would pass instead. Thus, at decision Node 4, an NLK Player B has an incentive to pass with a positive probability to mimic the level$_0$ player, which motivates an NLK Player A to pass with a positive probability at decision Node 5, as well. }

\begin{figure}[h]
    \centering
    \includegraphics[width=0.8\textwidth]{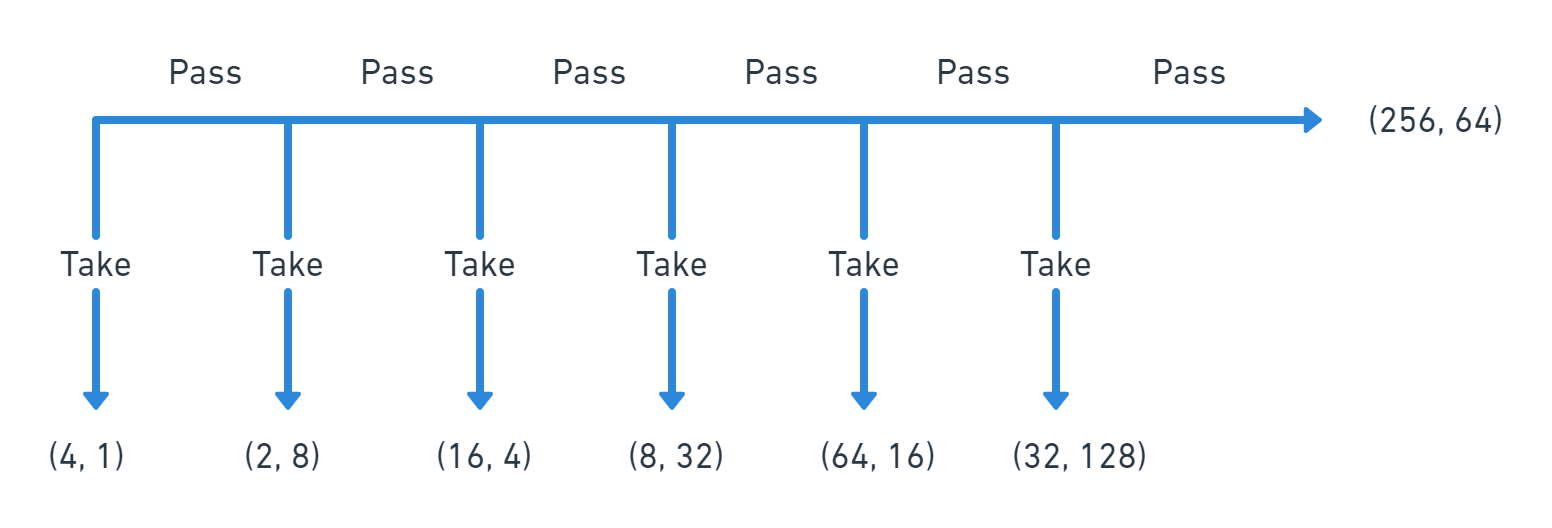}
    \caption{The Centipede game$^{34}$}
    \label{fig:1}
\end{figure}

We apply our model to experiment by~\citeN{palacios2009field} and~\citeN{levitt2011checkmate} on the \textit{centipede game}, where N$=3$ (Figure~\ref{fig:1}).

\footnotetext[34]{{This is the same example from~\citeN{palacios2009field}. Source: Drawn using \href{https://whimsical.com/}{https://whimsical.com/}.}}

The prediction of our PBNLK with all $\lambda \in\{0.05 n\}_{n=0,1,2, \ldots, 20}$ and the Level-K model with all $\mathrm{k} \in{\aleph}^{+}$and data from the aforementioned two papers are summarized in Table~\ref{tab:6} . When $\lambda=0$, PBNLK coincides with SPNE and level$_{k}, k \geq 6$, and when $\lambda \in[0.615,1]$, PBNLK coincides with level$_{1}$. For all other $\lambda \in(0,0.615)$, PBNLK generates different predictions. We first compare our predictions to data from Palacios-Huerta and Volij's laboratory experiment with four treatments. Unlike other experiments of the centipede game, in their work, the composition of two opponents varies\footnote[35]{See Table~\ref{tab:6} for details. The two opponents are chess players or students.} across treatments, and it is common knowledge among all players. This allows us to explore how beliefs represented by $\lambda$ and the results change as the nature of the subject pool changes. Next, we compare predictions| to data from \citeN{levitt2011checkmate}'s field experiments of chess players to further evaluate the predictions of NLK, because the data are different from the data of the former experiment.

{
\small
\renewcommand{\arraystretch}{1.2}
\begin{longtable}[h]{ccccccc}
\hline
\textit{Data or Prediction   }                               & \textit{Node 1} & \textit{Node 2} & \textit{Node 3} & \textit{Node 4} & \textit{Node 5} & \textit{Node 6} \\ \hline
\endfirsthead
\endhead
\begin{tabular}[c]{l@{}c@{}l}NLK & or & {} $level_k$ \\($\lambda$=0) & &  k $\geq 6$\end{tabular} & 1*              & 1               & 1               & 1               & 1               & 1               \\
\begin{tabular}[c]{@{}c@{}}NLK\\ ($\lambda$=0.05)\end{tabular}                                                    & 0               & 0.704           & 0.867           & 0.899           & 0.892           & 1               \\
\begin{tabular}[c]{@{}c@{}}NLK\\ ($\lambda$=0.1)\end{tabular}                                                     & 0               & 0.375           & 0.877           & 0.889           & 0.938           & 1               \\
    \begin{tabular}[c]{@{}c@{}}NLK\\ ($\lambda$=0.15)\end{tabular}                                                 & 0               & 0.007           & 0.889           & 0.889           & 0.999           & 1               \\
    \begin{tabular}[c]{@{}c@{}}NLK\\ ($\lambda$=0.2)\end{tabular}                                                 & 0               & 0               & 0               & 0.844           & 0.879           & 1               \\
\begin{tabular}[c]{@{}c@{}}NLK\\ ($\lambda$=0.25)\end{tabular}    & 0               & 0               & 0               & 0.792           & 0.887           & 1               \\
\begin{tabular}[c]{@{}c@{}}NLK\\ ($\lambda$=0.3)\end{tabular}                                                     & 0               & 0               & 0               & 0.732           & 0.895           & 1               \\
\begin{tabular}[c]{@{}c@{}}NLK\\ ($\lambda$=0.35)\end{tabular}                                                     & 0               & 0               & 0               & 0.663           & 0.905           & 1               \\
\begin{tabular}[c]{@{}c@{}}NLK\\ ($\lambda$=0.4)\end{tabular}                                                     & 0               & 0               & 0               & 0.583           & 0.916           & 1               \\
    \begin{tabular}[c]{@{}c@{}}NLK\\ ($\lambda$=0.45)\end{tabular}                                                 & 0               & 0               & 0               & 0.489           & 0.930           & 1               \\
\begin{tabular}[c]{@{}c@{}}NLK\\ ($\lambda$=0.5)\end{tabular}                                                     & 0               & 0               & 0               & 0.375           & 0.946           & 1               \\
\begin{tabular}[c]{@{}c@{}}NLK\\ ($\lambda$=0.55)\end{tabular}                                                     & 0               & 0               & 0               & 0.236           & 0.966           & 1               \\
\begin{tabular}[c]{@{}c@{}}NLK\\ ($\lambda$=0.6)\end{tabular}                                                     & 0               & 0               & 0               & 0.0625          & 0.991           & 1               \\
\begin{tabular}[c]{@{}c@{}}NLK \textit{or} {} $level_1$ \\ $(0.615 \leq \lambda \leq 1)$\end{tabular}                                            & 0               & 0               & 0               & 0               & 0               & 1               \\ 
        $level_2$   & 0               & 0               & 0               & 0               & 1               & 1               \\ 
        $level_3$                                            & 0               & 0               & 0               & 1               & 1               & 1               \\
                $level_4$                                    & 0               & 0               & 1               & 1               & 1               & 1               \\
                    $level_5$                                & 0               & 1               & 1               & 1               & 1               & 1               \\
Data$^{**}$ (S vs. S)  &
  \begin{tabular}[c]{@{}c@{}}0.030***\\ (200)\end{tabular} &
  \begin{tabular}[c]{@{}c@{}}0.17\\ (194)\end{tabular} &
  \begin{tabular}[c]{@{}c@{}}0.42\\ (161)\end{tabular} &
  \begin{tabular}[c]{@{}c@{}}0.65\\ (93)\end{tabular} &
  \begin{tabular}[c]{@{}c@{}}0.82\\ (33)\end{tabular} &
  \begin{tabular}[c]{@{}c@{}}0.83\\ (6)\end{tabular} \\
Data (S vs. C) &
\begin{tabular}[c]{@{}c@{}}0.30\\ (200)\end{tabular} &
  \begin{tabular}[c]{@{}c@{}}0.52\\ (140)\end{tabular} &
  \begin{tabular}[c]{@{}c@{}}0.61\\ (67)\end{tabular} &
  \begin{tabular}[c]{@{}c@{}}0.69\\ (26)\end{tabular} &
  \begin{tabular}[c]{@{}c@{}}1.00\\ (8)\end{tabular} &
  - \\
Data (C vs. S) &
  \begin{tabular}[c]{@{}c@{}}0.375\\ (200)\end{tabular} &
  \begin{tabular}[c]{@{}c@{}}0.44\\ (125)\end{tabular} &
  \begin{tabular}[c]{@{}c@{}}0.56\\ (70)\end{tabular} &
  \begin{tabular}[c]{@{}c@{}}0.61\\ (31)\end{tabular} &
  \begin{tabular}[c]{@{}c@{}}1.00\\ (12)\end{tabular} &
  \textbf{-} \\
Data (C vs. C) &
  \begin{tabular}[c]{@{}c@{}}0.725\\ (200)\end{tabular} &
  \begin{tabular}[c]{@{}c@{}}0.64\\ (55)\end{tabular} &
  \begin{tabular}[c]{@{}c@{}}0.90\\ (20)\end{tabular} &
  \begin{tabular}[c]{@{}c@{}}1.00\\ (2)\end{tabular} &
  - &
  - \\
Data$^{***}$ (Field) &
  \begin{tabular}[c]{@{}c@{}}0.039\\ (102)\end{tabular} &
  \begin{tabular}[c]{@{}c@{}}0.102\\ (98)\end{tabular} &
  \begin{tabular}[c]{@{}c@{}}0.193\\ (88)\end{tabular} &
  \begin{tabular}[c]{@{}c@{}}0.352\\ (71)\end{tabular} &
  \begin{tabular}[c]{@{}c@{}}0.587\\ (46)\end{tabular} &
  \begin{tabular}[c]{@{}c@{}}0.632\\ (19)\end{tabular} \\ \hline
\caption{Centipede game-Prediction and Data}
\label{tab:6}\\
\end{longtable}}

\vspace*{-20pt}
{\small
\noindent $^{*}$ presents the predicted probabilities of playing T at each node by the model. Columns correspond to the probability that a player is predicted to play T upon reaching that node. Odd nodes refer to Player A’s choices; even nodes refer to Player B’s choices.\\
$^{**}$ Data are from \citeN{palacios2009field}. S represents students, and C represents chess players. S vs. C represents the situation when Player A is a student and Player B is a chess player. C vs. S is when Player A is a chess player and Player B is a student. \\
$^{***}$ shows the distribution of implied stop probabilities for players in the centipede game. The number of opportunities observed is displayed in the parentheses. \\
$^{****}$ Data are from the field Centipede game of chess players by Levitt, List, and Sadoff (2011).
}

\vspace{6pt}

Referring to~\citeN{ho2013dynamic}, we define a measure, $D(H,M,G_S)$, to quantify the deviation of data H from the model’s prediction, $M$ in \textit{centipede game} $G_S$ with S decision nodes as follows:

\begin{equation*}
D\left(H, M, G_{S}\right)=\sum_{s=1}^{S} w_{s}^{H} d_{s}\left(p_{s}^{H}, p_{s}^{M}\right), w_{s}^{H}=\frac{n_{s}^{H}}{\sum_{k=1}^{S} n_{k}^{H}}, d_{s}\left(p_{s}^{H}, p_{s}^{M}\right)=\left|p_{s}^{H}-p_{s}^{M}\right|
\end{equation*}

where $n_s^H$ is the number of observations at each stage, given by data H, $d_s$ $(p_s^H,p_s^M)$ is the distance of stopping probabilities at stage s between data H and the prediction of model M measured by their absolute difference $|p_s^H-p_s^M|$.

{
\renewcommand{\arraystretch}{1.2}
\begin{longtable}{cccccc}
\hline Models &
  \begin{tabular}[c]{@{}c@{}}Data\\ (S vs. S)\end{tabular} &
  \begin{tabular}[c]{@{}c@{}}Data \\ (S vs. C)\end{tabular} &
  \begin{tabular}[c]{@{}c@{}}Data \\ (C vs. S)\end{tabular} &
  \begin{tabular}[c]{@{}c@{}}Data \\ (C vs. C)\end{tabular} &
  \begin{tabular}[c]{@{}c@{}}Data\\ (Field)\end{tabular} \\ \hline
\endfirsthead
\endhead
\begin{tabular}[c]{l@{}c@{}l}NLK & or & {} $level_k$ \\($\lambda$=0) & &  k $\geq 6$\end{tabular} & 0.7102   & 0.5474   & 0.5431   & 0.2773$^{\#*}$ & 0.7760   \\
\begin{tabular}[c]{@{}c@{}}NLK\\ ($\lambda$=0.05)\end{tabular} & 0.3016   & 0.2478  & 0.3191   & 0.5393      & 0.4296   \\
\begin{tabular}[c]{@{}c@{}}NLK\\ ($\lambda$=0.1)\end{tabular} & 0.2132   & 0.2361$^\#$ & 0.2619$^\#$ & 0.5785      & 0.3589   \\
\begin{tabular}[c]{@{}c@{}}NLK\\ ($\lambda$=0.15)\end{tabular} & 0.2071   & 0.3536   & 0.3672   & 0.6500      & 0.3269   \\
\begin{tabular}[c]{@{}c@{}}NLK\\ ($\lambda$=0.2)\end{tabular} & 0.1857   & 0.4051   & 0.4062   & 0.7166      & 0.2036   \\
\begin{tabular}[c]{@{}c@{}}NLK\\ ($\lambda$=0.25)\end{tabular} & 0.1791   & 0.4019   & 0.4023   & 0.7170      & 0.1946   \\
\begin{tabular}[c]{@{}c@{}}NLK\\ ($\lambda$=0.3)\end{tabular} & 0.1714   & 0.3982   & 0.3978   & 0.7175      & 0.1866   \\
\begin{tabular}[c]{@{}c@{}}NLK\\ ($\lambda$=0.35)\end{tabular} & 0.1625$^\#$ & 0.3971   & 0.3927   & 0.7181      & 0.1761   \\
\begin{tabular}[c]{@{}c@{}}NLK\\ ($\lambda$=0.4)\end{tabular} & 0.1703   & 0.4016   & 0.3905   & 0.7185      & 0.1638   \\
\begin{tabular}[c]{@{}c@{}}NLK\\ ($\lambda$=0.45)\end{tabular} & 0.1837   & 0.4069   & 0.3968   & 0.7192      & 0.1497   \\
\begin{tabular}[c]{@{}c@{}}NLK\\ ($\lambda$=0.5)\end{tabular} & 0.1999   & 0.4134   & 0.4044   & 0.7200      & 0.1323$^\#$ \\
\begin{tabular}[c]{@{}c@{}}NLK\\ ($\lambda$=0.55)\end{tabular} & 0.2197   & 0.4212   & 0.4137   & 0.7210      & 0.150    \\
\begin{tabular}[c]{@{}c@{}}NLK\\ ($\lambda$=0.6)\end{tabular} & 0.2444   & 0.4310   & 0.4253   & 0.7222      & 0.1818   \\
\begin{tabular}[c]{@{}c@{}}NLK \textit{or} $level_1$ \\ $(0.615 \leq \lambda \leq 1)$\end{tabular}   & 0.2840   & 0.4526   & 0.4569   & 0.7227      & 0.2121   \\
$level_2$ & 0.2533   & 0.4345   & 0.4295   & 0.7227      & 0.1933*  \\
$level_3$ & 0.2127*  & 0.4121   & 0.4139   & 0.7155      & 0.2428   \\
$level_4$ & 0.2502   & 0.3787*  & 0.3947*  & 0.6578      & 0.3703   \\
$level_5$ & 0.4366   & 0.3660   & 0.4290   & 0.6022      & 0.5540   \\
\begin{tabular}[c]{@{}c@{}} $level_k,k=1,2$ \\ \textit{(optimal distribution)} \end{tabular} & 0.2446   & 0.4345   & 0.4295   & 0.7227      & 0.1484   \\
\begin{tabular}[c]{@{}c@{}} $level_k,k=1,2,3$ \\ \textit{(optimal distribution)} \end{tabular} & 0.1567   & 0.3946   & 0.3872   & 0.7155      & 0.0895 \\ \hline
\caption{Centipede game-prediction for different models}
\label{tab:7}\\
\end{longtable}
}
\vspace*{-20pt}
{\small
* and \# indicate the best prediction of a single type Level-K and NLK, respectively. 
}

\vspace{6pt}

Table~\ref{tab:7} presents the result of $D(H,M,G_S)$ calculated using PBNLK and the Level-K model with the five different aforementioned data sets. In the lab experiments, when opponents are students (Column 2), PBNLK with $\lambda$=0.35 provides the most precise prediction (D=0.1625), which is better than the best prediction of the single type Level-K model (k=3,D=0.2127); in the treatment when chess players and students play with each other (Column 3 and 4), PBNLK with $\lambda$=0.1 fits the data the best $(D^{S vs C}=0.2361,D^{C vs S}=0.2619)$, which is more accurate than the Level-K model with an optimal k=4 $(D^{S vs C}=0.3787,D^{C vs S}=0.3947);$ when the opponents are chess players, the best fit is the case when PBNLK ($\lambda=0$), SPNE, and the $level_k$ type, ($k\geq 6)$, coincide ($D=0.2773$). For the field data, PBNLK with $\lambda$=0.5 provides the most precise prediction ($D=0.1323$), which is more accurate than the best prediction of the Level-K model with an optimal $k=2$  ($D=0.1933$). Moreover, in all five datasets, the optimal PBNLK performs better than the Level-K with an optimal distribution of level$_1$ and level$_2$ types. When we allow the Level-K model to have one more parameter, the optimal PBNLK still performs better the Level-K with an optimal distribution of level$_1$, level$_2$, and level$_3$ in three datasets (Column 2, 3, 4) except for the lab experiments when opponents are students and the field data. 

Our solution concept provides an alternative explanation for cases where neither the original Level-K model nor backward induction applies. Notably, we constrain NLK by using only symmetric beliefs. However, it is reasonable for each group to have a different subjective $\lambda$ in cases where students interact with chess players, and we conjecture that NLK would perform even better by allowing for heterogeneous beliefs while accounting for additional parameters. 

\section{Common Value Auction}

\citeN{avery1997second}(AK) conducted a laboratory experiment using a common value, second-price auction, and \textit{the wallet game}. Their design has two bidders, $i=1,2$, and each privately observes a signal $X_{i}$ drawn i.i.d from a \textit{uniform distribution} on $[1,4]$. The common value is the sum of the two private signals, i.e., $v_{i}\left(x_{1}, x_{2}\right)=v\left(x_{1}, x_{2}\right)=x_{1}+x_{2}$. Let $v(x, y)=x+y$, and $\mathrm{r}(\mathrm{x})=\mathrm{x}+\mathrm{E}\left[X_{2}\right]=x+2.5 . v(x, x) \equiv b(x)=2 x$ is the unique symmetric BNE\footnote[36]{Refer to \citeN{milgrom1982predation}.} and with just two bidders, $b(x)=2 x$, is an \textit{ex-post equilibrium}, independent of signals distribution and risk attitude and with no regret. AK defines \textit{naïve bidding} by $r(x)=x+2.5$, representing a naive bidder who assumes that whenever she wins, the other bidder's signal is at its expected value ( $2.5$ ). Notably, $r(x)$ is also the level$_{\text {player's strategy in Crawford and Iriberri }(\mathrm{CI} ; 2007), \text { the best }}$ response to a levelo player who bids uniformly randomly on $[1,4]$. We denote by $b^{\lambda}(\cdot)$ the strategy in a $\lambda-B N L K$ equilibrium and solve the symmetric linear strategy. (The details are provided in Appendix A.3.)

The data produced by $A K$ is evaluated using the cursed equilibrium (CE) model by~\citeN{eyster2005cursed}(ER) and the Level-K by~\cite{crawford2007level} (CI). ER show that for any cursed level, $0<\chi \leq 1$, their CE predicts better than BNE (i.e., CE with $\chi=0$ ) and that for a given $\chi, \mathrm{CE}$ fits better for experienced, rather than for inexperienced subjects, with respect to the MSE. For data on only inexperienced bidders, CI use Level-K with a logistic error structure and a subject-specific precision. They compare their model using the best mixture of five types, including random level$_1$ and level$_2$,\footnote[37]{Random level$_1$ and level$_2$ are generated iteratively by best responding to a random level$_0$, as considered in this paper.}  truthful level$_1$ and level$_2$,\footnote[38]{Truthful level$_1$ and level$_2$ are generated iteratively by best responding to a truthful level0 who always bids her signal: $b(x)=x$.}  and BNE players, and show that it outperforms CE (with the best mixture of types, such that $\chi\in{0.1,0.2,...,0.9,1.0})$, using both likelihood and the BIC.\footnote[39]{BIC penalizes models with more parameters to adjust the likelihood.}

Table~\ref{tab} compares the prediction of $\lambda$-BNLK (with all $\lambda \in\{0.05n\}_{n=0,1,2,..,20})$ and the Level-K model. The optimal bidding of a level2 player already reduces to a boundary solution (the objective function becomes a linear function), where all bidders with a value lower than 2.5 bid 3.5 and the others (with a value higher than 2.5) bid 6.5. For a level3 player, when her signal is smaller than 2.5, she bids any number below 3.5 (expecting to lose), and she bids any number above 6.5 when her signal is larger than 2.5 (expecting to win). The predictions are ambiguous for higher levels. By contrast, there always exists a symmetric linear strategy for our $\lambda$-BNLK players. 

{\renewcommand{\arraystretch}{1.2}
\begin{longtable}{cccc}
\hline
Models & b(x) & \begin{tabular}[c]{@{}c@{}}MSE\\ (inexperienced)\end{tabular} & \begin{tabular}[c]{@{}c@{}}MSE\\ (experienced)\end{tabular} \\\hline
\endfirsthead
\endhead
\begin{tabular}[c]{l@{}c@{}l}NLK & or & {} NE \\($\lambda$=0) & & \end{tabular} & 2x           & 2.897      & 1.171      \\
\begin{tabular}[c]{@{}c@{}}NLK\\ ($\lambda$=0.05)\end{tabular} & 1.951x+0.122 & 2.823      & 1.124      \\
\begin{tabular}[c]{@{}c@{}}NLK\\ ($\lambda$=0.1)\end{tabular} & 1.904x+0.239 & 2.756      & 1.082      \\
\begin{tabular}[c]{@{}c@{}}NLK\\ ($\lambda$=0.15)\end{tabular} & 1.859x+0.352 & 2.693      & 1.042      \\
\begin{tabular}[c]{@{}c@{}}NLK\\ ($\lambda$=0.2)\end{tabular} & 1.815x+0.462 & 2.634      & 1.010      \\
\begin{tabular}[c]{@{}c@{}}NLK\\ ($\lambda$=0.25)\end{tabular} & 1.772x+0.570 & 2.579      & 0.978      \\
\begin{tabular}[c]{@{}c@{}}NLK\\ ($\lambda$=0.3)\end{tabular} & 1.730x+0.676 & 2.531      & 0.953      \\
\begin{tabular}[c]{@{}c@{}}NLK\\ ($\lambda$=0.35)\end{tabular} & 1.688x+0.781 & 2.484      & 0.927      \\
\begin{tabular}[c]{@{}c@{}}NLK\\ ($\lambda$=0.4)\end{tabular} & 1.646x+0.886 & 2.440      & 0.906      \\
\begin{tabular}[c]{@{}c@{}}NLK\\ ($\lambda$=0.45)\end{tabular} & 1.604x+0.990 & 2.396      & 0.889      \\
\begin{tabular}[c]{@{}c@{}}NLK\\ ($\lambda$=0.5)\end{tabular} & 1.562x+1.096 & 2.356      & 0.872      \\
\begin{tabular}[c]{@{}c@{}}NLK\\ ($\lambda$=0.55)\end{tabular} & 1.519x+1.203 & 2.320      & 0.859      \\
\begin{tabular}[c]{@{}c@{}}NLK\\ ($\lambda$=0.6)\end{tabular} & 1.475x+1.313 & 2.286      & 0.848      \\
\begin{tabular}[c]{@{}c@{}}NLK\\ ($\lambda$=0.65)\end{tabular} & 1.430x+1.426 & 2.250      & 0.840      \\
\begin{tabular}[c]{@{}c@{}}NLK\\ ($\lambda$=0.7)\end{tabular} & 1.383x+1.543 & 2.220      & 0.835      \\
\begin{tabular}[c]{@{}c@{}}NLK\\ ($\lambda$=0.75)\end{tabular} & 1.333x+1.667 & 2.190      & 0.834*     \\
\begin{tabular}[c]{@{}c@{}}NLK\\ ($\lambda$=0.8)\end{tabular} & 1.281x+1.798 & 2.164      & 0.835      \\
\begin{tabular}[c]{@{}c@{}}NLK\\ ($\lambda$=0.85)\end{tabular} & 1.224x+1.940 & 2.137      & 0.843      \\
\begin{tabular}[c]{@{}c@{}}NLK\\ ($\lambda$=0.9)\end{tabular} & 1.161x+2.098 & 2.117      & 0.857      \\
\begin{tabular}[c]{@{}c@{}}NLK\\ ($\lambda$=0.95)\end{tabular} & 1.088x+2.280 & 2.097      & 0.882      \\
\begin{tabular}[c]{l@{}c@{}l}NLK & or & {} level$_1$ \\($\lambda$=1) & & \end{tabular} & x+2.5        & 2.085$^{\#*}$ & 0.922$^{\#*}$ \\
level$_2$ &   $\left\{\begin{array}{l}
3.5 \text { if } x<2.5 \\
6.5 \text { if } x>2.5
\end{array}\right.$           & 2.955      & 1.381      \\
level$_3$ & $\left\{\begin{array}{l}<3.5 \text { if } x<2.5 \\ >6.5 \text { if } x>2.5\end{array}\right.$             & -          &            \\
Data (inexperienced) &   $\begin{gathered}0.997 \\ (0.079)\end{gathered} {} \mathrm{x}+\begin{gathered}2.950 \\ (0.203)\end{gathered}$           & 1.899      &            \\
Data (experienced) &     $\begin{gathered}1.313 \\ (0.053)\end{gathered}{} \mathrm{x}+\begin{gathered}2.023 \\ (0.150)\end{gathered}$         & -          & 0.745     \\ \hline
\caption{Model comparison for the wallet game.}
\label{tab:8}\\
\end{longtable}
}

\begin{figure}[h]
    \centering
    \includegraphics[width=0.9\textwidth]{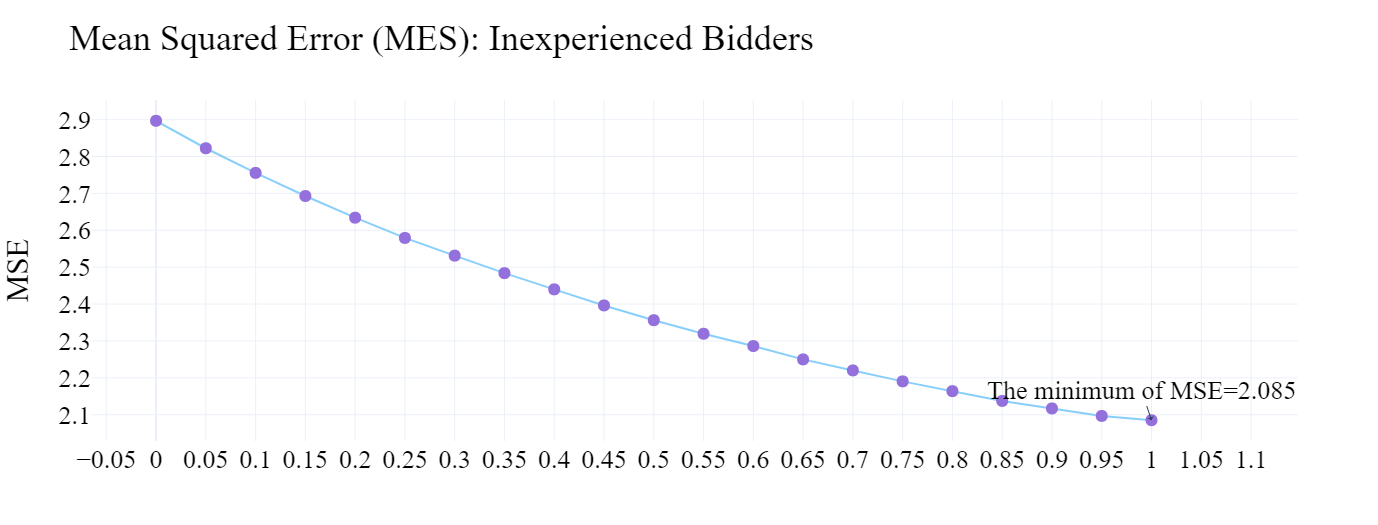}
    \caption{MSEs of BNLK with Different $\lambda$: inexperienced bidders.}
    \label{fig:2}
\end{figure}

\begin{figure}[!h]
    \centering
    \includegraphics[width=0.7\textwidth]{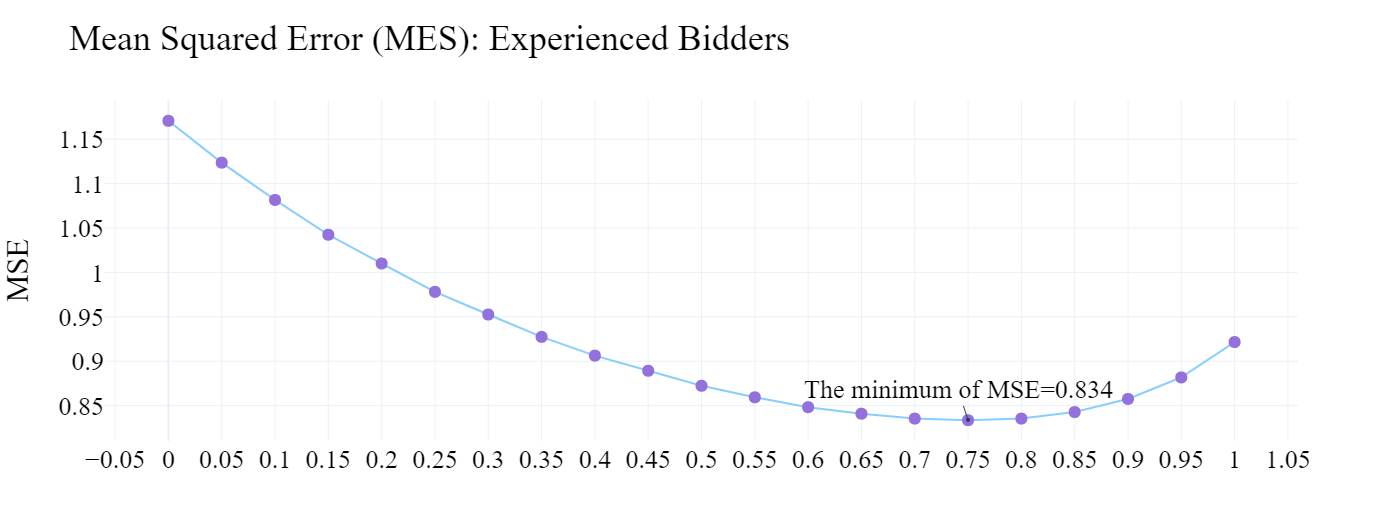}
    \caption{MSEs of BNLK with different $\lambda$: experienced bidders.}
    \label{fig:3}
\end{figure}

Table~\ref{tab:8} and Figures~\ref{fig:2} and~\ref{fig:3} show that for inexperienced bidders (using the first 18 periods), the most accurate prediction of BNLK is with $\lambda$=1, and it coincides with level1 (MSE=2.085).\footnote[40]{We choose the value of $\lambda$ that minimizes the mean squared errors (MSEs), that is, the nonlinear least squares estimate of $\lambda$.}  For experienced bidders (using periods 19-42), BNLK with $\lambda$=0.75 fits the data the best (MSE=0.834), which is better than the most precise prediction of Level-K (k=1,MSE=0.922). 

Thus far, we have used the in-sample of various versions of NLK, including BNLK and  PBNLK, to compare to other models.  However, in a recent paper,~\citeN{brunner2020self}, who studies experimental tournament games, compares the performance of PBNLK to that of the other models for \textit{out-of-sample} predictions. We cite his findings as follows: 

\begin{quote}
    Levin and Zhang (2019) have already shown that their PBNLK solution concept explains behavior in centipede games better than Nash equilibrium and Level-K thinking. this study finds that in a tournament context, PBNLK has the greatest out-of-sample predictive power in a modified version of the tournament among all the alternatives considered. Thus, PBNLK and its updating of beliefs during a game seem to be important concepts that promise to be valuable for the design of new policies and deserve more attention in future research.
\end{quote}

Finally, we looked at experimental data from the work of~\cite{gill2016cognitive} (G\&P) on the beauty contest game, and compared to the predictions of Nash equilibrium, Level-K and NLK.\footnote[41]{We thank a referee of the journal who suggested looking into the beauty contest game.} In this experiment there are three players, and guesses are from the set, I=\{0,1,2,...,100\}. The winner of a prize of \$6 (normalized here to 1) is the player whose guess, $g_i,i=1,2,3$, is the closest to 70\% of the target T, $T = \frac{7}{10}  \frac{g_1+g_2+g_3}{3}$, ties (of two or three) are equally divided.\footnote[42]{The main objective of their paper is to “study how cognitive ability measured in a nonstrategic setting affects how people perform and learn in a strategic environment.” They used a careful design and protocols to first classify exogenously experimental subjects to two groups of Low and High cognitive ability, and their treatments included playing 10 rounds of the game with the same three players that were of the same homogeneous group or a mixed of players from the two groups.} We used data only from the first round behavior,\footnote[43]{ ~\citeN{crawford2007level}, mid of page 7,  “In this paper we reconsider the winner's curse in common-value auctions and overbidding in independent-private-value auctions using a non-equilibrium \textbf{model of initial responses} based on "level-k" thinking, introduced by...” and bottom of page 7, “It also promises to establish a link between empirical auction studies and non-auction experiments on strategic thinking, and thereby to bring a large body of auction evidence to bear on the issue \textbf{of how best to model initial responses to games.}” Both bold added.}  and applied only the simplest versions of Level-K and NLK models. This is, in Level-K model, $level_0$ randomizes $uniformly$ on the set $I$ and player $level_{k+1}$ best responds by assuming all rivals are $level_k$ players, and the NLK player believes that each one (of the two rivals) is either a $level_0$ player, or an NLK player with probabilities $\lambda$, and (1-$\lambda$) respectively. We consider only symmetric NLK solutions.

\vspace{6pt}

\noindent \textbf{Case 1:} $\lambda$=0: In this case the NLK player faces two other NLK players, thus the solution coincides with Nash equilibrium, and the unique \textit{dominance solvable} outcome is: $g_2^*=0$. 

\vspace{6pt}

\noindent \textbf{Case 2:} $\lambda=1:$ In this case the NLK player faces two level$_{0}$ players and the NLK model is identical to the Level-K model, where the, $N L K$, or level$_{1}$, players maximize the probability of winning. (For now we ignore the integer constraint and solve as if on uniform $[0,1]$ ).
For $g<\frac{7}{16}$, the probability of NLK winning with a guess of $g$ is given by:\footnote[44]{For $\frac{7}{8} \leq g$ and for $\frac{7}{16} \leq g<\frac{7}{8}$, the probability of the DM winning with $g$ are given by
$(1-g)^{2}$, and by $(1-g)^{2}+\frac{1}{56}(7-8 g)^{2}$, respectively and both strictly decline in $g$.} $(1-g)^{2}-$ $\frac{1}{56}(7-16 g)^{2}+\frac{8}{7} g^{2}+\frac{2}{7}\left(7 g-16 g^{2}\right)=(1-g)^{2}-\frac{1}{56}(7-16 g)^{2}+\frac{2}{7}(7-12 g) g$, with first-order-condition for maximization being $2[2-7 g]=0$, yielding a guess of $g^{*}=\frac{2}{7}=$ $0.2857$ as being (optimal) the best response for level$_{1}$ and $\mathrm{NLK}$, and with $\frac{273}{392}=0.6964$, being the probability of winning. In the simple version of Level-K the best response of a level$_{2}$ player facing two level$_{1}$ rivals is any guess below $g^{*}$.

The average first round guesses in G\&P among players who were classified as having High and Low cognitive abilities are $42.9$ and $43.6$ respectively. Neither the NLK, nor the Level-K models predict subjects' behavior well, with Nash equilibrium being much worse.\footnote[45]{From their figure~\ref{fig:2}. On page 1632 it seems that the corresponding average guesses for rounds 6 -10 is about 10.4 and 12.6 for High and Low cognitive abilities, respectively. } 

\vspace{6pt}

\noindent \textbf{Case 3:}  $0<\lambda<1$:  Ignoring the constraint imposed by restricting guesses only to integers we show (omitted) that an NLK player would undercut any presumed strictly positive, g$>$0 solution, thus eliminating the possibility of a pure strategy solution. Yet, the reason is telling: the tiniest undercutting from a presumed g$>$0 solution induces a positive, discontinuous increase in payoffs of $\frac{2}{3}$, when the NLK player is matched with two other NLK players, (as the player wins 1 rather than $\frac{1}{3}$ by a using the same guess), and a gaining 1 rather than $\frac{1}{2}$ when the NLK player is matched with one NLK player and one $level_0$ player whose guess is larger than the presumed $g>0$.  On the other hand, when the NLK is matched with two $level_0$ players, undercutting hurts, as a lower guess is even further from the optimal guess of 0.2857. However, because that loss is “continuous,” (at 0.2857 it vanishes), there is $\epsilon>0$ undercutting, small enough, to validate such intuition. In G\&P paper, guesses are restricted to be integers and cannot be as small as we wish and the loss is not infinitesimal and we can always find small enough $\delta>0$, such that $\lambda’s$ satisfying, $1-\delta<\lambda<1$, will rationalize (say) a guess of 28 as a pure strategy NLK solution.\footnote[46]{It suggests that in a beauty contest game with guesses allowed only from the set, $I=\{0,10,20,...,100\}$, a guess of 30 can be “supported” as a pure strategy NLK for a “larger” interval of $\lambda<1$.  Such design may be better suited for comparing Level-K and NLK without sacrificing much of the original motivation of that game. }  

Solving NLK’s mix strategy equilibrium for each $\lambda$, $0<\lambda<1$, is a computational nightmare and we have not done it. We conjecture it has the following qualitative property.\footnote[47]{We made enough computation to have fate in this conjecture.}   For a given $\lambda$ there is an interval, [$l_\lambda,0.2857]$, from which NLK players pick their guess using endogenously determined distribution. The mix strategy equilibrium implies that a tiny undercutting induces continuous change that thus must be “balanced” against moving further from the optimal NLK guess in the event where the NLK player is  matched with two $level_0$ players. Qualitatively, it results in predictions that resembles the behavior of $level_2$, player, but provides a precise mixing whereas the Level-K model allows any guess under below $g^*$. 

We conclude that with $0<\lambda<1$, NLK does better than $level_1$, $level_2$, that predicts a guess of 0.2857, or 0 respectively, but we do not call it a winner as it uses an extra parameter.\footnote[48]{  If we allow a $level_2$ player in the Level-K model best respond to a mix of rivals who are $level_1$ and $level_0$, we will have similar qualitative predictions.}  

\section{Conclusion}

This introduces NLK, a model that connects NE and Level-K. NLK allows a player to believe that her opponent may be less or as sophisticated as herself.  NLK is well-defined in both static and dynamic games, making it easy to apply to the data from four published papers on static, dynamic, and auction games. In all four cases, NLK provides better predictions than those of the NE and Level-K, except for few cases when predictions coincide or when we allow Level-K to freely choose more parameters
NLK allows other beliefs about the naïve player and can be extended to manage heterogeneous beliefs about opponents (e.g., who are from distinct populations) and to games with more than two players.

Comparing across applications to experimental data from different games with different cognitive requirements of the tasks, we observe that the “best fitting” $\lambda$ depends on the game played and the population of players. Based on our intuition and the limited evidence, ceteris paribus, the expectation should be a smaller $\lambda$ in simpler games requiring less cognitive/strategical sophistication, or with more sophisticated/experienced (e.g., chess) players. For instance, in the common value auction with inexperienced bidders, $\lambda$=1 provides the best fit, but with experienced bidders, $\lambda$=0.75 fits better.  We also find that in the centipede game, the best $\lambda s$ are smaller than those in the common value auction. This finding suggests that further experimental research is necessary with the same, or similar, games with similar populations, to estimate the optimal $\lambda$, to provide more evidence and tests for external validity.\footnotetext[49]{For instance, Brünner (2018) documents the consistently good performance of NLK in two similar games with the same $\lambda$: a regular rank-order tournament and the version with an outside option.} 
\citep{alaoui2016endogenous}

\newpage

\nociteth{*}
\bibliographystyleth{ACM-Reference-Format}
\bibliographyth{NLK}

\newpage

\appendix

\section{Appendix A}

\subsection{Solve for $\lambda$-NLK equilibrium in the Money Request Game}

We only go through the solution for $0 \leq \lambda<\frac{1}{2}$ because a similar argument follows for $\frac{1}{2} \leq \lambda<1$.

We first claim that when $0 \leq \lambda<\frac{1}{2}, \$ 20$ must be played by an $\mathrm{NLK}$ player. We assume for contraction that $\$ 20$ will not be played; then, deviation to $\$ 20$ would end up with $\$ 20$ for sure, whereas choosing $\$ 19$ generates $\$ 19+\lambda \times \frac{1}{10} \times 20(<20)$. Thus, $\$ 19$ will not be played by an NLK player. By induction, no strategy is valid for an NLK player. This is a contradiction.\footnote[50]{Because the game is finite, by Proposition 1, NLK exists. } Thus, $\$ 20$ must be played by an NLK player. However, $\$ 20$ could not be the only pure strategy of an NLK player because he has an incentive to deviate to $\$ 19$. We assume that $j<19$ is the largest number played with positive probability. Hence, deviating to $\$ 19$ generates a strictly larger payoff. Then, $\$ 19$ must be played with positive probability. We denote the probability of playing $\$$ j in the NLK equilibrium by $\beta_{j}, \beta_{j} \in[0,1]$ and $\sum \beta_{j}=1$. The expected payoff of all strategies in equilibrium should be the same, and because playing $\$ 20$ yields $\$ 20$ for sure, it follows that $19+(1-\lambda) \beta_{20} 20+\lambda \frac{20}{10}=20 .$ Then, $\beta_{20}^{*}=\frac{1-2 \lambda}{(1-\lambda) 20}<1 .$ By the same argument, $18+(1-\lambda) \beta_{19} 20+\lambda \frac{20}{10}=20 .$ Then, $\beta_{19}^{*}=\frac{2-2 \lambda}{(1-\lambda) 20}$
Because $\beta_{20}^{*}+\beta_{19}^{*}<1, \$ 18$ has to be played in equilibrium (otherwise, there would be an incentive to deviate to $\$ 18) ;$ thus, iteratively, we obtain $\beta_{18}^{*}=\frac{3-2 \lambda}{(1-\lambda) 20}, \beta_{17}^{*}=\frac{4-2 \lambda}{(1-\lambda) 20}, \beta_{16}^{*}=$ $\frac{5-2 \lambda}{(1-\lambda) 20} .$ We suppose that $\$ 14$ is played in equilibrium too; then, $14+(1-\lambda) \beta_{15} 20+\lambda \frac{20}{10}=20$ implies that $\beta_{15}=\frac{6-2 \lambda}{(1-\lambda) 20}$. However, in this case, $\sum_{j=15}^{20} \beta_{j}>1$. This is a contradiction. Thus, $\$ 14$ (and all lower numbers) would not be played by an NLK player. Then, $\beta_{15}^{*}=1-\sum_{j=16}^{20} \beta_{j}=$ $\frac{5-10 \lambda}{(1-\lambda) 20} .$ In conclusion, when $0 \leq \lambda<\frac{1}{2}$, there is a unique mixed strategy for an $\mathrm{NLK}$ player where $\left\{\sigma_{i}^{*}\right\}=\left\{\beta_{15}^{*}, \beta_{16}^{*}, \beta_{17}^{*}, \beta_{18}^{*}, \beta_{19}^{*}, \beta_{20}^{*}\right\}=\left\{\frac{5-10 \lambda}{(1-\lambda) 20}, \frac{5-2 \lambda}{(1-\lambda) 20}, \frac{4-2 \lambda}{(1-\lambda) 20}, \frac{3-2 \lambda}{(1-\lambda) 20}, \frac{2-2 \lambda}{(1-\lambda) 20}, \frac{1-2 \lambda}{(1-\lambda) 20}\right\} .$

\subsection{Solve for $\lambda$-NLK equilibrium in the common value auction }

We assume there is a linear pure strategy for a $\lambda$-NLK player and denote it as $b^{\lambda}(x)=$
$b^{\lambda}(1)+\frac{b^{\lambda}(4)-b^{\lambda}(1)}{3}(x-1), x \in[1,4] .$ We denote $d^{\lambda}=b^{\lambda}(4)-b^{\lambda}(1) .$ The probability that the opponent is level $_{0}$ conditional on a tie is $q^{\lambda}=\operatorname{Pr}\left(\right.$ rival $=$ level $_{0} \mid$ tie at bid $\left.=b\right)=$ $\frac{\lambda / 6}{\lambda / 6+(1-\lambda) / d^{\lambda}}, b \in\left[b^{\lambda}(1), b^{\lambda}(4)\right] \subseteq[2,8] .$ Then, by indifference, in the case of the maximum willingness to pay conditional on a tie, we denote it by $M W P(X)=b(x)$ :

$
\begin{aligned}
&M W P(1)=q^{\lambda}(1+2.5)+\left(1-q^{\lambda}\right) 2=1.5 q^{\lambda}+2=b^{\lambda}(1) \\
&M W P(4)=q^{\lambda}(4+2.5)+\left(1-q^{\lambda}\right) 8=8-1.5 q^{\lambda}=b^{\lambda}(4) \\
&\text { Then, } d^{\lambda}=b^{\lambda}(4)-b^{\lambda}(1)=1-3 q^{\lambda}=\frac{\lambda / 6}{\lambda / 6+(1-\lambda) / d^{\lambda}} \\
&\text { Then, }\left(d^{\lambda}\right)^{2}+3 \frac{2-3 \lambda}{\lambda} d^{\lambda}-\frac{36(1-\lambda)}{\lambda}=0
\end{aligned}
$

Thus, the bidding strategy is $b^{\lambda}(x)=b^{\lambda}(1)+\frac{d^{\lambda}}{3}(x-1)$, where $b^{\lambda}(1)=1.5 q^{\lambda}+2$, $d^{\lambda}=\frac{3}{2 \lambda}\left(3 \lambda+\sqrt{-7 \lambda^{2}+4 \lambda+4}-2\right)$, and $q^{\lambda}=\left(1-d^{\lambda}\right) / 3$

\subsection{Comparisons with Other Related Solution Concepts }
~\citeN{eyster2005cursed} propose CE that also relaxes the restriction on beliefs in NE, while maintaining the equilibrium concepts for players' strategies. They show that CE rationalizes behavior (data) from experiments where BNE fails. In particular. this rationalization occurs in common value auctions, where~\cite{kagel2002information} observe systematic overbidding and losses, a phenomenon called the winner's curse. $\mathrm{CE}$ also fits experimental data from voting and signaling models better than $\mathrm{BNE}$. In one extreme version of $\mathrm{CE}$, entitled "fully $\mathrm{CE}$," individuals correctly predict other players' distribution of actions but ignore the correlation between actions and the specific players' types who chose those actions. In their general model, $\chi$-CE, beliefs are a weighted average of beliefs in fully cursed opponents (with weight $\chi$ ) and Bayesian Nash opponents (with weight $(1-\chi)$ ). The CE characterizes heterogeneous behaviors by different cursed levels (with $\chi=1$ being fully cursed, and $\chi=0$ being BNE). However, CE reduces to NE when there is complete information. Hence, it cannot be applied to explain deviations from NE in both static and dynamic games with complete information. Conceptually, the CE models bounded rational agents as only partially taking.into.account how other players' actions depend on their type. By contrast, NLK allows a player to consider the possibility that the other player is a naive player or another NLK player like herself. \cite{kets2012bounded} extends the type space of~\cite{harsanyi1973games}, by allowing players to have a finite instead of infinite depth of reasoning. However, different from NLK, it required that the depth of reasoning be the same for all players, and beliefs that the other player might be less sophisticated are not allowed.
\par
For applications to dynamic games with perfect information and recall, analogy-based expectation equilibrium (ABEE), a solution concept proposed by~\citeN{jehiel2005analogy}, is the most closely related to ours.\footnote[51]{~\citeN{jehiel2008revisiting}) extends his analogy-based concept to Bayesian games.} In $\mathrm{ABEE}$, agents first group the set of opponents' decision nodes into a partition, namely, an analogy class. Next, they form expectations of each opponent's average behavior at every element of the analogy class rather than, more precisely, at each decision node. Although conceptually, $\mathrm{ABEE}$ is similar to $\mathrm{CE}$, when applied to a different type of games, $\mathrm{ABEE}$ also suggests that individuals might not fully consider how others' choices depend on their information, and such deficiency in reasoning is common knowledge among all players.\footnote[52]{More specifically, information means the history upon reaching a decision node at which the choice is made.} By contrast, our model allows NLK players to consider heterogeneity in their opponents' inference process. In our adaptation of NLK equilibrium to dynamic games, beliefs about different types of opponents are
\par
anchored at the beginning of the game and are updated at each stage using Bayes' rule. Analytically, ABEE coincides with SPNE for the finest analogy partition, and similar to NLK, ABEE can also rationalize passing, in the centipede game, to the last few stages for a large range of partitions, in violation of the backward induction predictions. However, ABEE does not provide a specific means to choose an analogy class, whereas NLK offers a means of parametric estimation to specify beliefs in equilibrium.\footnote[53]{As an extension of the Level-K model to dynamic games, \citeN{ho2013dynamic} apply their model to the experiment data of the centipede game. However, they intend to study learning across repetitions, whereas ours explains strategic behavior better even for novel games. Moreover, unlike their model, NLK does not restrict the strategy set, which allows NLK to capture Bayesian updating for beliefs across stages within one round.} 

As with all of the aforementioned solution concepts, NLK maintains the best response to beliefs but relaxes NE's requirement of a player's consistent beliefs about other players. In dynamic games, \cite{aumann1992irrationality}, similar to several other writers in the subsequent literature, has shown that a failure of backward induction does not imply a failure of individual rationality. For example, in the centipede game, backward induction implies that the first mover ought to use stop at the first decision node, which has rarely been demonstrated in experimental data. These papers show that some relaxations of the "common knowledge of rationality" explain several rounds of passing, although all of the players are individually rational. \footnote[54]{An individual must be careful about the terminology according to the epistemic condition of NE. \citeN{aumann1995epistemic} prove that in a two-person game, mutual knowledge of preferences and payoffs, rationality, and beliefs regarding the other players’ strategies are sufficient for NE. This common knowledge of rationality is unnecessary for NE in a two-person game. Moreover, \citeN{battigalli1999recent} argue that there is a contradiction between the results of backward induction and a common belief in sequential rationality at later stages. Thus, in this paper, by “common knowledge of rationality,” we mean, in general, the extra assumptions necessary for NE/BNE/SPNE other than individual rationality.}$^,$\footnote[55]{\citeN{aumann1992irrationality} shows that continuation of the game beyond the first node for several rounds could occur even with “mutual knowledge” of high degrees. Considering that some sequentially rational behaviors off the equilibrium path are only reachable by the violation of sequential rationality, \citeN{reny1992rationality} defines a weaker version of sequential rationality in light of forward induction. \citeN{ben1997rationality} proves that cooperation in the centipede game is consistent with the common certainty of rationality, a weaker concept than the common knowledge of rationality. \cite{Asheim2003} introduce the concept of “fully permissible sets” to the extensive form game, where players reason deductively by attempting to determine one another’s moves. They show that deductive reasoning does not necessarily imply backward induction.} 

\end{document}